\documentstyle[11pt,aaspp4]{article} 

\lefthead{Yoshida, Lee}
\righthead{}

\newcommand{\half}{{\case{1}{2}}}
\newcommand{\be}{\begin{equation}}
\newcommand{\ee}{\end{equation}}
\newcommand{\ba}{\begin{eqnarray}}
\newcommand{\ea}{\end{eqnarray}}
\newcommand{\da}{\dagger}

\begin{document}


\title{Inertial modes of slowly rotating isentropic stars}
\author{Shijun Yoshida and Umin Lee}
\affil{Astronomical Institute, Tohoku University, Sendai 980-8578, 
Japan \\ yoshida@astr.tohoku.ac.jp, lee@astr.tohoku.ac.jp}

\begin{abstract}

We investigate inertial mode oscillations of slowly and uniformly
rotating, isentropic, Newtonian stars. Inertial mode oscillations are 
induced by the Coriolis force due to the star's rotation, and their 
characteristic frequencies are comparable with the rotation frequency 
$\Omega$ of the star. So called $r$-mode oscillations form a sub-class 
of the inertial modes. In this paper, we use the term ``$r$-modes'' to 
denote the inertial modes for which the toroidal motion dominates the 
spheroidal motion, and the term ``inertial modes'' to denote the inertial 
modes for which the toroidal and spheroidal motions have comparable 
amplitude to each other. Using the slow rotation approximation consistent 
up to the order of $\Omega^3$, we study the properties of the inertial 
modes and $r$-modes, by taking account of the effect of the rotational 
deformation of the equilibrium on the eigenfrequencies and eigenfunctions.
The eigenfrequencies of the $r$-modes and inertial modes calculated in 
this paper
are in excellent agreement with those obtained by Lindblom et al (1999) and
Lockitch \& Friedman (1998).
We also estimate the dissipation timescales due to 
the gravitational radiation and several viscous processes for
polytropic neutron star models. 
We find that for the inertial modes, the mass multipole gravitational 
radiation dominates the current multipole radiation, which is
dominating in the case of the $r$-modes.  
It is also found that the inertial mode instability is more 
unstable than previously reported by Lockitch \& Friedman (1998), and 
survives the viscous damping processes relevant in neutron stars.

\end{abstract}



\section{Introduction}

Since Andersson (1998) suggested and Friedman \& Morsink (1998) analytically
 verified
that the $r$-modes of rotating stars are
unstable due to the gravitational radiation reaction,
much attention has been paid to oscillation modes of rotating neutron stars
because of their possible importance in a field of astrophysics.
Lindblom et al. (1998) argued that due to
this instability, the maximum angular rotation velocity of hot young neutron
stars is strongly restricted.
Owen et al. (1998) also suggested that
the gravitational radiation emitted from hot young neutron stars due to
the $r$-mode instability is expected to be one of the potential sources
for the gravitational wave detectors, e.g. LIGO. 
Following these earlier investigations on the $r$-mode instability, there 
have appeared a number of 
papers which study the properties of the $r$-modes of 
rotating stars (e.g., Kojima 1998, Anderson et al 1998, Kokkotas \& 
Stergioulas 1998, Lindblom \& Isper 1998, Beyer \& Kokkotas 
1999, Lindblom et al 1999, Kojima \& Hosonuma 1999).

It is well known that the $r$-modes form a sub-class of the modes called  
``inertial modes'', for which
the restoring force is the Coriolis force and the characteristic frequency
is comparable to the angular rotation frequency $\Omega$ of the star 
(e.g., Greenspan 1964).
As the prominent characteristics of the $r$-modes, we know that the toroidal
motion
dominates the spheroidal motion in the velocity field, and that
the frequency $\omega$ observed in the corotating frame of the star
is given by $\omega=2m\Omega/(l(l+1))$ in the limit of $\Omega\rightarrow0$,
where $m$ and $l$ are the indices of a spherical harmonic function $Y_l^m$ 
representing the toroidal component of the velocity field 
(e.g., Papaloizou \& Pringle 1978 , Provost et al 1981, Saio 1982).
As shown by Lee et al (1992), however, the inertial modes in general
have the spheroidal component of the velocity field comparable with 
the toroidal component and do not have any analytical formula to
give $\omega/\Omega$ in the limit of 
$\Omega\rightarrow0$, except for the case of the Maclaurin spheroids
(Lindblom \& Ipser 1998; see also Bryan 1889).
In this paper, to distinguish these two classes of inertial modes,
we use the term of ``inertial modes'' to refer to the inertial
modes with the comparable spheroidal and toroidal components of the velocity
 field, and
the term of ``$r$-modes'' to refer to the inertial modes whose
toroidal component of the velocity field is dominating the spheroidal 
component.

Very recently, Lockitch \& Friedman (1998) calculated the inertial modes, 
assuming an ordering law given by
\be
\begin{array}{rrrr}
S_l \sim O(1),      & H_l   \sim O(1), & iT_{l'} \sim O(1), & \\
\delta \rho \sim O(\Omega^2), & \delta p \sim O(\Omega^2), & 
\delta \Phi \sim O(\Omega^2), & \sigma \sim O(\Omega),             
\label{ordering}
\label{order}\end{array}
\ee
where $S_l$ and $H_l$ are the spheroidal components and $iT_{l'}$ is the 
toroidal component of the displacement vector, and $\delta\rho$, $\delta p$,
and $\delta \Phi$ are the Eulerian perturbations of the density, pressure, 
and gravitational potential, and $\sigma$ is the oscillation frequency 
observed in an inertial frame. For $r$-modes, we may assume an ordering law 
given as $\sigma\sim O(\Omega)$, $iT_{l'}\sim O(1)$, and 
$S_l\sim H_l\sim \delta\rho\sim \delta p\sim \delta\Phi\sim O(\Omega^2)$.
Lockitch \& Friedman (1998) showed that a number of inertial modes 
are unstable because of the gravitational radiation reaction, just as 
$r$-modes are, and estimated the dissipation timescales due to the 
gravitational radiation and 
some viscous processes for slowly rotating polytropic neutron star models, 
where they made use of eigenfrequencies and eigenfunctions for 
slowly rotating Maclaurin spheroids to classify the inertial modes. 
Notice that Lindblom \& Ipser (1998) obtained analytically eigenfrequencies 
and eigenfunctions of the inertial mode for the Maclaurin spheroids 
(see also Bryan 1889).
Lockitch \& Friedman (1998) and Lindblom \& Ipser (1998) called 
the inertial modes ``rotation modes'' or ``generalized $r$-modes'', 
which are exactly the same as 
the inertial modes that will be discussed in this paper.

In this paper, 
we study the inertial modes and $r$-modes of slowly rotating stars, by 
employing a different
numerical approach to the problem (Lee \& Saio 1986, Lee 1993) 
from the calculations by, e.g.,
Lindblom et al. (1999), and Lockitch \& Friedman (1998). 
Newtonian gravity, uniform rotation, and isentropic structure of the 
equilibrium 
are assumed for adiabatic perturbations in this paper.
The effects of the rotational deformation of the equilibrium structure on the
eigenfunctions and eigenfrequencies are included in the analysis, where
the formulation by Lee (1993) (see also Lee \& Saio 1986) has been improved 
so that
the eigenmodes are consistent up to the order of $\Omega^3$.
The plan of this paper is as follows. 
In \S 2, we describe briefly the basic equations for the equilibrium 
configurations of slowly rotating polytropes. 
In \S 3, we present 
the formalism to calculate normal modes of slowly rotating stars.  
In \S 4, the properties of the eigenfrequencies and eigenfunctions of the 
inertial modes are discussed.
In \S 5, we examine the stability of simple neutron star models
against the inertial modes and the $r$-modes, computing their growth or 
damping timescales due 
to the gravitational radiation reaction and some viscous damping processes.
\S 6 is for discussions and conclusions.


\section{Equilibrium State}

We consider oscillations of an uniformly rotating, isentropic, Newtonian 
star.  
The structure of an equilibrium state is determined by the hydrostatic 
equation, the Poisson equation, and the equation of state:
\be
\nabla_i \, p = - \rho \, \nabla_i \Psi \, , \label{hy_st_eq}
\ee
\be
\nabla_i \nabla^i \, \Phi = 4 \pi G \rho \, , \label{Poi_eq}
\ee
\be
p = p (\rho) \, , \label{EOS}
\ee
where $\Psi$ is the effective potential defined by 
\be
\Psi =  \Phi - {1 \over 3} \, \Omega^2 r^2 \, 
       \{1 - P_2(\cos \theta)\}  \, , \label{def_pot}
\ee
and $\Omega$ is the angular rotation frequency of the star,
constant for uniform rotation.
Here we use spherical polar coordinates $(r,\theta,\phi)$, and 
$P_l(\cos \theta)$ denotes the Legendre polynomial, and $\nabla_i$ is 
the covariant derivative with respect to the quantity followed.

In this investigation, assuming slow 
rotation, we apply the Chandrasekhar-Milne expansion 
(see, e.g, Tassoul 1978) to equations (\ref{hy_st_eq})--(\ref{def_pot}). 
In this technique the effects of the centrifugal 
force and the equilibrium deformation are treated 
as small perturbations to a non-rotating spherically symmetric star. 
The small expansion parameter due to rotation is chosen as 
$\bar{\Omega}=\Omega (R^3/G M)^{1/2}$, where $R$ and $M$ are 
the radius and the mass of the non-rotating star, respectively. 
To the lowest order effects of the centrifugal force, the effective potential 
$\Psi$ can be expanded as follows:
\be
\Psi(r,\theta) = \Psi_0 (r) -2 R^2 \Omega^2 \,
[\psi_0 (r/R) + A_2 \psi_2 (r/R) P_2(\cos \theta) + O(\bar{\Omega}^2)] 
\, , \label{def_pot2}
\ee
where $\Psi_0 (r)$ is the potential of the non-rotating star.  
The functions $\psi_0(x)$ and $\psi_2(x)$ are determined by solving 
the following ordinary differential equations:
\be
\frac{1}{x^2} \frac{d}{dx} \left( x^2 \, \frac{d\psi_0}{dx} \right)
= k(x) \psi_0 (x) + 1 \, , \label{psi0}
\ee
\be
\frac{1}{x^2} \frac{d}{dx} \left( x^2 \, \frac{d\psi_2}{dx} \right)
= \left( k(x) + \frac{6}{x^2} \right) \psi_2 (x) \, , \label{psi2}
\ee
\be
k(x) = 4 \pi G R^2 \frac{d \rho_0}{d \Psi_0} \, ,
\ee
where $x= r/R$ and $\rho_0$ is the density of the non-rotating star.
For simplicity, we consider a sequence of slowly rotating stars whose 
central density is the same as that of the non-rotating star.  
From this condition, together with the regularity condition at the center 
of the star,  
the boundary conditions of equations (\ref{psi0}) and (\ref{psi2}) at 
the origin are given as 
\be
\psi_i(0)=0 \, , \hspace{.5in} \frac{d \psi_i(0)}{d x} = 0 \, ,
\ee 
where $i=0,\ 2$. 
From the boundary conditions at the stellar surface, the constant $A_2$ is 
determined as 
\be
A_2 = - {5 \over 6 \,
\left( 3 \psi_2 (1) + d \psi_2(1)/ d \ln x \right) } \, .
\ee
%


\section{Perturbation Equations}

In the perturbation analysis, we introduce
a parameter $a$ that is constant on a distorted effective potential 
surface. In practice, the parameter $a$ is defined such that
\be
\Psi (r,\theta) = \Psi_0 (a) \, . \label{ep-co}
\ee
With this parameter $a$, the distorted equi-potential surface may be given 
by
\be
r = a \lbrace 1 + \epsilon (a,\theta,\phi) \rbrace \, .
\label{def-a}
\ee
By using equations (\ref{def_pot2}), (\ref{ep-co}) and (\ref{def-a}), 
we obtain the explicit 
expression for the function $\epsilon (a,\theta) $ up to the order 
of $\bar{\Omega}^2$:
\be
\epsilon (a,\theta) = \alpha ( a ) + \beta (a) P_2 (\cos \theta ) \, ,
\ee
where
\be
\alpha (a) = \frac{2 c_1 \bar{\Omega}^2 \psi_0(x)}{x^2} \, ,
\ee
\be
\beta (a) = \frac{2 c_1 \bar{\Omega}^2 A_2 \psi_2(x)}{x^2} \, .
\ee
Here, $c_1 = (a/R)^3/(M(a)/M)$, and $M(a)$ denotes the mass inside the 
$a$-constant surface.

Hereafter, we employ the parameter $a$ as the radial coordinate, instead of 
the polar radius coordinate $r$. 
In this coordinate system $(a,\theta,\varphi)$, the metric tensor is written
by  
\be
ds^2 = ( 1 + 2 \epsilon ) 
       ( da^2 + a^2 d \theta^2 +a^2 \sin^2 \theta d \varphi^2) 
     + 2 a \epsilon_{,a} da^2 + 2 a \epsilon_{,\theta} da d \theta 
     + O (\bar{\Omega}^4)   \, , \label{metric}
\ee
where a comma denotes a partial derivative with respect to the 
variables followed. 
Note that in this frame the  pressure, density 
and effective potential of a rotating star depend only on the one variable $
a$,
although the orthogonality of the basis vectors is lost.

The governing equations of nonradial oscillations of a rotating star are 
obtained by
linearizing the basic equations used in fluid mechanics.
Since the equilibrium state is assumed to be stationary and axisymmetric, 
the perturbations may be
represented by a Fourier component proportional to 
$e^{i(\sigma t + m \varphi)}$, where $\sigma$ is the frequency observed in 
an inertial frame and $m$ is the azimuthal quantum number. 
The continuity equation may be linearized to be
\be
\delta \rho = - \nabla_i (\rho \xi^i), \label{del_rho} \label{mass_con} 
\ee
where $\xi^i$ is the Lagrangian displacement vector, and we have made use of
$\delta v^i  = i(\sigma + m \Omega) \xi^i \equiv i \omega \xi^i$
with $\omega$ being the oscillation frequency observed in the corotating 
frame of the star.  The linearized Euler's equation is
\be
-\omega^2 \xi_i + \nabla_i 
\left( \frac{\delta p}{\rho} + \delta \Phi \right) + 
A_i \frac{\delta p}{\rho} +\xi^j A_j \frac{1}{\rho} \nabla_i p +
2 i \omega \Omega \xi^j \nabla_j \varphi_i 
= 0 \, , \label{pert_Euler}
\ee
where $\varphi^i$ is the rotational Killing vector, with which the 3-velocity
of the equilibrium fluid of a rotating star is given as $v^i=\Omega\varphi^i
$.
The last term on the left-hand-side of equation ($\ref{pert_Euler}$)
comes from the Coriolis force, and its explicit components in the metric 
($\ref{metric}$) are given by 
\be
\xi^j \nabla_j \varphi_a = - (1+2 \epsilon +a \epsilon_{,a}) \, 
a \sin^2 \theta \xi^\varphi + O (\bar{\Omega}^4) \, ,
\ee
\be
\xi^j \nabla_j \varphi_\theta = - \lbrace (1+2 \epsilon) \, 
\cos \theta + \epsilon_{,\theta} \sin \theta \, \rbrace \, 
a^2 \sin \theta \xi^\varphi + O (\bar{\Omega}^4) \, ,
\ee
\be
\xi^j \nabla_j \varphi_\varphi = - (1+2 \epsilon+a \epsilon_{,a}) \,
a \sin^2 \theta \xi^a - \lbrace (1+2 \epsilon) \, \cos \theta + 
\epsilon_{,\theta} \sin \theta \, \rbrace \, 
a^2 \sin \theta \xi^\theta + O (\bar{\Omega}^4) \, .
\ee
(Note that Lee (1993) ignored the terms of $\epsilon$, $\epsilon_{,a}$, and
$\epsilon_{,\theta}$ in $\xi^j\nabla_j\varphi_i$, as a result of which 
the eigenmodes obtained are consistent only up to the order of $\bar\Omega^2
$.
By retaining these terms in the analysis, the eigenmodes calculated in this 
paper are consistent up to the order of $\bar\Omega^3$ for slow rotation.)
The perturbed Poisson equation is given by
\be
\nabla_i \nabla^i \delta \Phi = 4\pi G\delta \rho \, .
\label{pert_Poisson}
\ee
For adiabatic oscillations, we have
\be
\delta p    = \Gamma p \left( \frac{\delta \rho}{\rho} +
\xi^i A_i \right)  \,   \label{del_p} 
\ee
where $\Gamma$ is the adiabatic index given by
\be
\Gamma = \left( \frac{ \partial \ln p}{\partial \ln \rho} \right)_{ad} \, ,
\ee
and $A_i$ is the generalized Schwarzschild discriminant defined by
\be
A_i = \frac{1}{\rho} \nabla_i \rho - \frac{1}{\Gamma p} \nabla_i p \, .
\label{def_A}
\ee
In the case of isentropic stars, the 1-form $A_i$ vanishes exactly.

Physically acceptable solutions of the linear differential equations are 
obtained by imposing boundary conditions at the inner and outer boundaries of 
the star.
The inner boundary conditions are the regularity condition of the perturbed 
quantities 
at the stellar center. 
The outer boundary conditions at the surface of the star are 
$\Delta p/\rho=0$
and the continuity of the perturbed gravitational potential at the surface to
the solution of $\nabla_i\nabla^i\delta\Phi=0$ which vanishes at infinity.

In order to solve the system of partial differential equations given above, 
we employ a series expansion in terms of spherical harmonics to
represent the angular dependence of the perturbed quantities.
The Lagrangian displacement vector, $\xi^i$, is expanded in terms of the 
vector spherical harmonics as
\be 
\xi^a = \sum_{l\geq\vert m \vert}^{\infty} a \, S_l(a)  
Y_l^m (\theta,\varphi) e^{i \sigma t} \, ,   \label{v_a}
\ee
\be
\xi^\theta = \sum_{l,l'\geq\vert m \vert}^{\infty} 
\left\{ H_l (a) {Y_l^m}_{,\theta}  + T_{l'} (a) \frac{1}{\sin \theta} \, 
{Y_{l'}^m}_{,\varphi} \right\} e^{i \sigma t} \, ,     \label{v_t}
\ee
\be\xi^\varphi = \frac{1}{\sin^2 \theta} 
\sum_{l,l'\geq\vert m \vert}^{\infty} \left\{ H_l (a) {Y_l^m}_{,\varphi}
        - T_{l'} (a) {\sin \theta} \, {Y_{l'}^m}_{,\theta}
        \right\} e^{i \sigma t} \,   \label{v_p} 
\ee
(Regge \& Wheeler 1957, see also Thorne 1980).
The perturbed scalar quantity such as $\delta \Phi$ is expressed as
\be 
\delta \Phi  = \sum_{l\geq\vert m \vert}^{\infty} \delta \Phi_l (a) \, 
Y_l^m (\theta,\varphi) e^{i \sigma t} \, .  \label{d_phi}
\ee
Substituting the perturbations into the linearized basic equations 
($\ref{del_rho}$), ($\ref{pert_Euler}$), ($\ref{pert_Poisson}$) 
and ($\ref{del_p}$),
we obtain an infinite system of coupled linear ordinary 
differential
equations for the expansion coefficients. The details are given in the 
Appendix.

For numerical calculations,
we truncate the infinite set of linear ordinary differential equations to 
obtain a finite set by discarding all the expansion coefficients
associated with $l$ higher than $l_{max}$. 
This truncation determines the number of the expansion coefficients 
kept in the spherical harmonic expansion of each perturbed quantity. 
We denote this number as $k_{max}$.
Note that the number $k_{max}$ is equivalent to the dimension of the column 
vectors
${\bf y}_i$, introduced in the Appendix, for $i=1$-$4$.
Our basic equation therefore becomes a 
system of $4\times k_{max}$-th order ordinary differential equations, which,
together with the boundary conditions,
are numerically solved as an eigenvalue problem of $\omega\equiv \sigma+m\Omega$ 
by using a Henyey type relaxation method (e.g., Unno et al . 1989, see also
Press et al. 1992).

For the number $k_{\mbox{\tiny max}}$ of the expansion 
coefficients kept in the eigenfunctions, we employ 
$k_{\mbox{\tiny max}}$=5 or 6, values that are found to be 
large enough for the inertial modes calculated in this paper.
Note that if we consider higher radial-order modes, with many radial nodes 
of the eigenfunctions, than the modes obtained in this study, 
the effects of truncation can be more serious, as shown by Lee et al. 
(1992).

When $\Omega=0$, nonradial oscillation modes are classified into two decoupled 
sets, called
the ``polar'' parity mode (or the ``spheroidal'' mode) and 
the ``axial'' parity mode (or the ``toroidal'' mode). 
Although these modes have been traditionally called 
``even'' and ``odd'' parity modes, respectively, we will use these terms 
for different meanings in this paper (see below). 
When $\Omega\not=0$, both parity modes are mixed, and
an oscillation mode contains contributions from the polar and axial parity 
modes.

Since the equilibrium state of a rotating star is invariant under the parity
transformation defined by $\theta \to \pi - \theta$, and
is symmetric with respect to the equator, the linear perturbations have  
definite parity for that transformation.  In this paper,
a set of modes whose scalar perturbations such as $\delta \Phi$ are
symmetric with respect to the equator is called ``even'' modes, while
a set of modes whose scalar perturbations are antisymmetric with respect to 
the equator
is called ``odd'' modes (see, e.g., Berthomiue et al. 1978).
For positive integers $k=1,~2,~3,~\cdots$, we have $l=|m|+2k-2$ and $l'=l+1$
for even modes, and $l=|m|+2k-1$ and $l'=l-1$ for odd modes, 
where the symbols $l$ and $l'$ have been used
to denote the spheroidal components and toroidal components of the 
displacement vector $\xi^i$, respectively 
(see equations (27)$-$(30)).
Notice that in Lockitch \& Friedman (1998), the terms ``even'' and 
``odd'' 
modes are used to denote the ``polar-led hybrids'' and ``axial-led hybrids''
 modes, respectively.


\section{The Eigenvalues and Eigenfunctions of Inertial Modes}

We compute the eigenvalues and eigenfunctions of 
inertial modes and $r$-modes of rotating polytropic stars 
with accuracy up to the order of $\bar{\Omega}^3$. 
In this study, we concentrate our attention to the case of isentropic stars,
for which the generalized Schwarzschild discriminant $A_i$ vanishes exactly 
and
the equilibrium state is marginally stable against convection. 
In this case, the adiabatic index is 
given by using the polytropic index $n$ as
\be \Gamma = \frac{d \ln p}{d \ln \rho} = \frac{n + 1}{n} \, .  
\ee
Since the rotational deformation is of the order of $\bar\Omega^2$,
the effects on the frequencies ($\propto\bar\Omega$) 
of the inertial modes and $r$-modes 
appear as terms of the order of $\bar\Omega^3$.
For sufficiently small values of $\bar{\Omega}$, therefore,
we may expand the eigenfrequency $\omega$ of the inertial modes and $r$-modes
of a rotating star in powers of $\bar\Omega$ as 
\be
\frac{\omega}{\Omega} =  \kappa_0 + \kappa_2 \, \bar{\Omega}^2 
 + O(\bar{\Omega}^4)  \, . \label{def-kappa}
\ee
In this study, the expansion coefficients $\kappa_0$ and $\kappa_2$ are 
obtained
by fitting the eigenfrequencies computed for a mode at several small values 
of
$\bar\Omega$.

For nonradial $p$-, $f$-, and $g$-modes 
of a non-rotating or sufficiently slowly rotating star, 
we may specify the values of $l$ and $m$
without any ambiguity and, by counting the number of radial nodes of the 
eigenfunctions, 
we may order the modes without any confusion (e.g., Unno et al. 1989).
But, to our knowledge, no well-established classification scheme for inertial 
modes of a rotating star
exists, except for the case of the Maclaurin spheroids, for which
exact eigensolutions are known (Bryan 1889; Lindblom \& Ipser 1998).
As shown by Bryan (1889), eigenmodes of the Maclaurin spheroid can be 
characterized by the eigenvalue $\omega/\Omega$ and 
the angular quantum numbers $l_0$ and $m$ of the associated
Legendre functions of the spheroidal coordinate defined on a surface form 
of the spheroid. 
The functional forms of the eigenfunctions in the spheroidal coordinate,
which depends on the eigenvalue $\omega/\Omega$,
are specified by $l_0$ and $m$. 
Since the the spheroidal coordinate employed is dependent on $\omega/\Omega$,
the functional forms in the actual spatial coordinate are different for  
different eigenvalues $\omega/\Omega$ for given $l_0$ and $m$. In fact, 
Lindblom \& Ipser (1998) have shown that, for given $l_0$ and $m$, 
the eigenvalue $\kappa_0$ for inertial modes of the Maclaurin spheroid
can be determined, to lowest order of $\bar\Omega$, 
by solving the algebraic equation given by
\be
m \, \frac{d^m}{d x^m}  P_{l_0}(\kappa_0 /2)
+ \left( \frac{\kappa_0}{2} - 1 \right) \, 
 \frac{d^{m+1} }{d x^{m+1}} P_{l_0}(\kappa_0 /2) = 0 \, , \label{eigen-eq}
\ee
where $m\ge0$ is assumed. They also have shown that
\be
-2 < \kappa_0 < 2 \, ,  \label{fre-ina}
\ee
and that the number of the roots of equation (\ref{eigen-eq}) is equal to 
$l_0-m$.

In this paper, as a classification scheme for inertial modes of uniformly 
rotating polytropic stars with $n\not=0$, 
we follow the scheme used for the inertial modes of the Maclaurin 
spheroid,
and we pick up the inertial modes that are similar in the mode character to 
those of the Maclaurin spheroid, labeling them with the quantum numbers $l_0
$ and $m$.
In practice, the value of $l_0$ is 
determined by examining the properties of the eigenfunctions, 
such as the number of radial nodes of 
the expansion coefficients, and the number of the
dominant expansion coefficients of the eigenfunctions. 
A similar procedure was employed by Lockitch \& Friedman (1998). 
Notice that our definition of $l_0$ is not the same as that of Lockitch \& 
Friedman (1998), but
the same as that of Lindblom \& Ipser (1998).

In Table 1 we tabulate the eigenvalues $(\kappa_0, \kappa_2)$ for 
inertial modes and $r$-modes of a polytropic model with $n=1$. 
As expected from the results for the Maclaurin spheroid, 
the value of $l_0 -  |m|$ is odd for odd modes and even for
even modes. 
The modes which satisfy the condition $\sigma(\sigma+m\Omega)<0$ are
unstable to the gravitational radiation reaction (see the next section), and
are marked with an asterisk $\ast$. 
We find that a number of inertial modes are unstable to 
the gravitational radiation reaction, as first shown by Lockitch \& Friedman
 (1998).  
The modes with $l_0 - \vert m \vert = 1$ are $r$-modes and the frequency
to lowest order in $\Omega$ is given by 
$\kappa_0=2/(\vert m \vert + 1)$,
which is independent of the structure of the star.
Only the $r$-modes with $l_0 - 1=\vert m \vert$, for which 
the toroidal component $iT_{l'}$ associated with $Y_{l'=|m|}^m$ is dominating,
are found in isentropic stars, as suggested by Saio (1982).
Table 1 also shows that, although $r$-modes are all retrograde, 
there are both prograde and retrograde inertial modes.

In Table 2, the eigenfrequencies $(\kappa_0, \kappa_2)$ 
of inertial modes and $r$-modes with $m=2$
are tabulated for several values of the polytropic index $n$, to see the 
dependence of the
eigenfrequencies on the equation of state, where 
the eigenfrequencies for the case of $n=0$ 
are obtained by using the results by Lindblom \& Ipser (1998).  
As shown by Table 2, the eigenvalues ($\kappa_0$,$\kappa_2$) of the inertial
 modes depend on
the polytropic index $n$, while only the eigenvalue $\kappa_2$ of the 
$r$-modes is dependent on the index $n$.

Let us compare our calculations to those by 
Lindblom et al (1999) and by Lockitch \& Friedman (1998).
We find that
the $r$-mode eigenfrequencies $(\kappa_0,\kappa_2)$ given in Table 1 for $m$
=2 and 3
are in good agreement with those given in table 1 of Lindblom
et al. (1999).  
Note that the normalization applied to $\kappa_2$ 
employed by Lindblom et al. (1999) is different, by a factor $4/3$, 
from that employed in this paper.
We also find that the inertial mode frequencies $\kappa_0$ given in Tables 1
 and 2 are
in good agreement with those given in the table 6 of Lockitch \& Friedman 
(1998),
who do not give any number for $\kappa_2$ because of their neglect of
the effects of the rotational deformation.

In Figures 1 to 3, we show the several expansion coefficients 
$S_l(a/R)$, $H_l(a/R)$ and ${\it i} T_{l'}(a/R)$
for four different $m=2$ inertial modes of a polytropic model with $n=1$,
where the amplitude normalization is given at the surface of the model by
$iT_{|m|+1}=1$ for even modes and $iT_{|m|}=1$ for odd modes.
In Figure 1, the first two expansion coefficients are plotted, versus the 
fractional radial coordinate defined as $a/R$, for the even parity inertial 
modes with $m=2$ and $l_0-\vert m \vert=2$. The solid curves give 
the expansion coefficients corresponding to the mode with $\kappa_0=1.100$, 
while the dotted curves correspond to the mode having $\kappa_0=-0.557$.
This figure shows that only the first expansion coefficients are 
dominating for both axial and polar components of the displacement vector. 
In Figures 2 and 3, the first three expansion coefficients of 
the spheroidal and toroidal components of the displacement vector are 
plotted against the fractional radial coordinate $a/R$ for the unstable 
even parity inertial modes with $m=2$ and $l_0-\vert m \vert=4$. 
Figures 2 and 3 correspond to the mode with $\kappa_0=1.520$ and 
with $\kappa_0=0.863$, respectively. The figures show that only the first 
two expansion coefficients of the eigenfunctions have dominant amplitude.
Figures 1 to 3 may confirm that the series expansion of the eigenfunctions 
in terms of vector spherical harmonics converges quickly for the inertial 
modes calculated here. We note that the behavior of the eigenfunctions is 
very similar to that of the corresponding modes of the Maclaurin spheroid
(see also Lockitch \& Friedman 1998).
If we consider the modes associated with given $l_0$ and $m$,
the number of nodes of the dominant expansion coefficients 
in the radial direction is the same for the modes with different 
eigenvalues $\kappa_0$. This means that the number of radial nodes of the 
eigenfunctions is not necessarily a good quantum number to label the modes 
with different eigenvalues $\kappa_0$ but with the same $l_0$ and $m$.

Figures 1 to 3 also show that for inertial modes of a slowly rotating 
isentropic star, the toroidal expansion coefficients $i T_{l'}$ has 
comparable amplitude to the spheroidal expansion coefficients $S_l$ and 
$H_l$ in the interior of the star.
It is also shown that since the eigenfunctions $S_l$ have vanishing amplitude 
at the surface, the fluid motion is almost tangential 
to the stellar surface.

In Table 3, we tabulate the number of radial nodes of the dominant expansion
coefficients of $\xi^i$ of the $r$-modes and inertial modes for several values 
of $l_0-|m|$. Note that the content of Table 3 is the same for $m=1,~2,~3$. 
The number of the dominant expansion coefficients 
of the eigenfunctions $iT_{l'}$ may be given as the maximum positive 
integer $k$ that satisfies $2k-1\leq l_0-|m|$. Also the number of the 
dominant expansion coefficients of the spheroidal components, $S_l$ and 
$H_l$, may be given as $k$ for the even parity modes and as $k-1$ for the 
odd parity modes.  Although we cannot give any 
proof in general, Table 3 may show that the number of radial nodes of the 
dominant expansion coefficients $i T_{l'}$ with the highest harmonic index 
$l'$ is zero, and that the number of radial nodes of the dominant expansion 
coefficients $i T_{l'}$ with the lowest harmonic index $l'$ is given by the 
maximum positive integer $k$ which satisfies $2k+1\leq l_0-|m|$.


\section{Dissipation Timescales}

As mentioned in the previous section, some of the inertial modes calculated
in this paper are unstable to the gravitational radiation reaction in the 
sense that
the frequency $\sigma$ satisfies the condition $\sigma(\sigma+m\Omega)<0$
(see, equations (\ref{E-D}) and (\ref{E-J})).
However, since in neutron stars 
there are some possible viscous and dissipative processes which tend to
suppress the instability, to decide whether the rotating neutron star is 
really unstable due to
the gravitational radiation reaction,
we need to know that the instability is strong enough to survive the
dissipative processes.

The effects of the gravitational radiation reaction and viscous processes on
the $r$-modes 
have already been studied by a number of authors (Lindblom et al 1998, 
Owen et al 1998, Andersson et al 1998, Kokkotas \& Stergioulas 1998, 
Lindblom et al 1999), and they have found that for the $r$-modes
the instability due to the gravitational radiation reaction
strongly dominates the viscous damping processes considered.  
For inertial modes, Lockitch \& Friedman (1999) suggested that
the inertial mode instability due to the gravitational radiation survives
the dissipative processes considered for the case of $r$-modes, but 
the instability itself is weaker than that of the $r$-modes.
In this section, we reconsider these estimations using a different numerical
approach to compute the oscillation modes of a rotating star.

To estimate the dissipation timescales associated with viscosity and 
gravitational radiation reaction, we employ a simple method used for 
the analysis of the $r$-mode instability (Lindblom et al 1998, see also 
Ipser \& Lindblom 1991). 
When the dissipation timescale is sufficiently long 
compared to the oscillation period of the mode, the growth rate or the
damping rate is approximately evaluated with the non-dissipative 
eigenfunctions as 
\be
\frac{1}{\tau} = - \frac{1}{2E} \frac{dE}{dt},   \label{tau}
\ee
where $E$ is the energy of the oscillation, observed in the corotating 
frame, given by 
\be
E = \frac{1}{2} \int \left[ 
\rho \delta v^i \delta v^{\ast}_i 
+ \left( \frac{\delta p}{\rho}+\delta\Phi\right) \delta\rho^{\ast}
\right] d^3 x \, ,  \label{E}
\ee
and the asterisk $^*$ denotes the complex conjugate of the indicated 
quantity.
The time derivative of $E$ is determined by the dissipation effects, 
for which we consider the shear viscosity, bulk viscosity and 
gravitational radiation.

The dissipation rate due to the shear viscosity can be calculated from 
\be
\left( \frac{dE}{dt} \right)_S = - 2 \, \int
\eta\delta\sigma^{ij}\delta\sigma_{ij}^{\ast} \, d^3 x \, ,
\label{E-S}
\ee
where $\delta\sigma_{ij}$ is the shear of the perturbation written by 
\be
\delta\sigma_{ij} = \half\left( \nabla_i \delta v_j + 
\nabla_j \delta v_i - \case{2}{3} g_{ij} \nabla_k \delta v^k \right) \, ,
\ee
and the coefficient of shear viscosity for hot neutron star matter is given 
by
\be
\eta = 2\times 10^{18}
\left(\frac{\rho}{10^{15}\mbox{g}\!\cdot\!\mbox{cm}^{-3}}
\right)^{\case{9}{4}}
\left(\frac{10^9K}{T}\right)^2 \
\mbox{g}\!\cdot\!\mbox{cm}^{-1}\!\cdot\!\mbox{s}^{-1} \,    \label{eta}
\ee
(Cutler \& Lindblom 1987; Sawyer 1989).
The dissipation rate due to the bulk viscosity can be written as 
\be
\left( \frac{dE}{dt} \right)_B = - \int
\zeta \delta\theta\delta\theta^{\ast} \, d^3 x \, , \label{E-B}
\ee
where $\delta\theta$ is the expansion of the perturbation defined as
\be
\delta\theta \equiv \nabla_i \delta v^i \, 
 = - \frac{{\it i} \omega}{\Gamma p}  \, \left( 
 \delta p + \xi^a \frac{dp}{da} \right)  \, ,
\ee
and the bulk viscosity coefficient for hot neutron star matter is
\be
\zeta = 6\times 10^{25}
\left(\frac{1\mbox{Hz}}{\sigma + m\Omega}\right)^2
\left(\frac{\rho}{10^{15}\mbox{g}\!\cdot\!\mbox{cm}^{-3}}\right)^2
\left(\frac{T}{10^9K}\right)^6 \
\mbox{g}\!\cdot\!\mbox{cm}^{-1}\!\cdot\!\mbox{s}^{-1} \,   \label{zeta}
\ee
(Cutler \& Lindblom 1987; Sawyer 1989).

The dissipation due to gravitational radiation comes from two divisible 
contributions, which are
\be
\left( \frac{dE}{dt} \right)_{G-D} = -\sigma (\sigma + m\Omega) 
\sum_{l=2}^{\infty} N_l \sigma^{2l} \left|\delta D_{lm}\right|^2 \, , 
\label{E-D}
\ee
and 
\be
\left( \frac{dE}{dt} \right)_{G-J} = -\sigma (\sigma + m\Omega) 
\sum_{l=2}^{\infty} N_l \sigma^{2l} \left|\delta J_{lm}\right|^2 \, ,
\label{E-J}
\ee 
where the coupling constant $N_l$ is given by 
\be
N_l = \frac{4\pi G}{c^{2l+1}}\frac{(l+1)(l+2)}{l(l-1)[(2l+1)!!]^2}.
\ee
Here, the mass, $\delta D_{lm}$, and current, $\delta J_{lm}$, multipole 
moments of the perturbation are given by (Thorne 1980, Lindblom et al. 1998)
\be
\delta D_{lm} = \int \delta\rho r^l Y_l^{\ast m} d^3 x \, ,  \label{D}
\ee 
and
\be
\delta J_{lm} = \frac{2}{c} \left(\frac{l}{l+1}\right)^{\half} \int r^l 
\left( \rho\delta v_i + \delta\rho v_i \right) Y^{i,B\ast}_{lm} d^3 x  
\, ,  \label{J}
\ee
and $Y^{i,B}_{lm}$ is the magnetic type vector spherical harmonic given by
\be
Y^{i,B}_{lm} = \frac{r}{\sqrt{l(l+1)}} \epsilon^{ijk} \nabla_j Y_l^{m} 
\nabla_j r
\ee
(Thorne 1980).
Notice that as indicated by equation (\ref{J}), the gravitational radiation 
due to 
the current multipole moment is emitted by the ``axial'' 
parity perturbations.

For inertial modes, since the first term in equation (\ref{E}) is dominant 
and 
the energy $E$ of the modes is positive definite for sufficiently small 
$\Omega$, 
the instability sets in when $dE/dt>0$.
As shown by equations (\ref{E-D}) and (\ref{E-J}), since
only the energy change rate due to gravitational radiation can be positive 
and the energy change rates due to the other dissipative processes are 
all negative, 
the necessary condition for the instability is that there exist modes 
whose frequencies satisfy $\sigma(\sigma+m\Omega)<0$ and
the right-hand-side of equations (\ref{E-D}) and (\ref{E-J}) becomes 
positive. 
Physically, this means that the rotating star becomes
unstable when the gravitational radiation carries away negative energy.

We may write the damping timescale of the mode
to the lowest order in $\bar{\Omega}$ as follows:
\begin{eqnarray}
\frac{1}{\tau} &=& \frac{1}{\tilde \tau_S} \left( \frac{10^9 K}{T} \right)^2
+ \frac{1}{\tilde \tau_B} \left( \frac{T}{10^9 K} \right)^6
\left( \frac{\pi G \bar{\rho}}{\Omega^2} \right) \nonumber \\
&+& \sum_{l=2}^{\infty} \frac{1}{\tilde \tau_{J,l}} 
\left( \frac{\Omega^2}{\pi G \bar{\rho}} \right)^{l+1}
+ \sum_{l=2}^{\infty} \frac{1}{\tilde \tau_{D,l}} 
\left( \frac{\Omega^2}{\pi G \bar{\rho}} \right)^{l+2} \, ,
 \label{tau2}
\end{eqnarray}
where $\bar{\rho}$ is the average density of the star. Here, the first, 
second, 
third and fourth terms in equation (\ref{tau2}) are contributions from 
the shear viscosity, the bulk viscosity, the current multipole radiation and
the mass multipole radiation, respectively. 
The expression of the timescale $\tau$ given by equation (\ref{tau2}) is 
basically the same as
that given by Lockitch \& Friedman (1999), who
ignored the last term, considering that the gravitational radiation
from the mass multipole moment is negligible for inertial modes.
However, for the even parity modes, the largest components from 
the mass multipole moment and the current multipole moment may be $\delta 
D_{\vert m \vert , m}$ and $\delta J_{\vert m \vert + 1 , m}$, respectively.
If both of the components have the same amplitude, the radiation from 
the mass multipole moment is by a factor $\bar{\Omega}^{-2}$ larger than 
that of 
the current multipole moment. 
Since the square of the mass multipole moment is by a factor 
$\bar{\Omega}^2$ 
smaller than that of the current multipole moment because of the ordering law
$\delta \rho\sim O(\bar\Omega^2)$ and 
$\delta v^i\propto\omega\xi^i\sim O(\bar\Omega)$
given by equation (1), 
the radiation reaction from the mass multipole is  
the same order as that from the current multipole for the even parity 
inertial modes. 
For consistency, therefore we must include the mass 
multipole moment as well as the current multipole moment to estimate the 
energy change rate.

In Table 4, we tabulate the timescales in 
the unit of second for 
various dissipative processes for a polytropic model with the index 
$n=1$,
where the radius and the mass at $\Omega=0$ are chosen to be
$R=12.57\mbox{km}$ and $M=1.4M_{\sun}$, respectively.  
As shown by Table 4, for inertial modes the mass multipole 
radiation dominates the current multipole radiation for both the even and odd 
parity modes.
Note that for $r$-modes with $l_0-|m|=1$, 
only the current multipole moment is relevant to the
gravitational radiation.
We also find that for even parity inertial modes, 
the fastest growth time due to the mass multipole 
radiation is of the order of $10^3$ seconds 
for the parameters $10^9K$ and $\Omega=\sqrt{\pi G\bar{\rho}}$.

Most of the timescales calculated in this paper are in good agreement
with those obtained 
by Lockitch \& Friedman (1998) and Lindblom et al. (1999), and
the relative difference between the calculations are at most several 
percents. 
However, there are a few discrepancies between our calculation and the
calculations by
Lockitch \& Friedman (1998) and Lindblom et al. (1999).
We note that the most dominating current multipole $\delta J_{lm}$ is 
that associated with $l_{max}$
and the moment $\delta J_{lm}$ with $l<l_{max}$ is less important than 
$\delta J_{l_{max}m}$,
where $l_{max}$ is the largest value of $l$ associated with the 
dominating expansion coefficients of $\delta v^a$. 
The result that for the odd parity inertial modes the multipole moment 
$\delta J_{lm}$ with $l<l_{max}$ does not vanish may contradict the result 
obtained by Lockitch \& Friedman (1998), who suggested that 
for the odd parity inertial modes all the current multipole moments 
$\delta J_{lm}$ with $l<l_{max}$ vanish (or nearly vanish).
The origin of these discrepancies is not clear at the moment.

As seen from equation (\ref{tau2}), the total dissipation timescale 
$\tau$ is given as a function of the angular velocity $\Omega$ and the 
temperature $T$ of a star. 
For a given temperature $T$, we can define the critical angular velocity 
$\Omega_c$ such that
\ba
\frac{1}{\tau (\Omega_c , T)} = 0  \hspace{.3in} 
\mbox{for} \hspace{.3in}  0 < \Omega_c < (\pi G \bar{\rho})^{1/2} \, .
\label{def-oc}
\ea
With this critical $\Omega_c$, we may say that a rotating neutron star is 
unstable
due to the gravitational radiation when $\Omega > \Omega_c$ because
the gravitational radiation reaction dominates the viscous damping 
processes. 
In Figure \ref{fig5}, the critical angular rotation velocities $\Omega_c$ for 
two inertial modes and an $r$-mode are plotted against the 
temperature of the star. 
From this figure we can see that the instability 
of the $m=2$ even parity inertial mode with $\kappa_0=1.100$ is 
strong, although this instability is weaker than that of the $r$-mode.   
At about $T=10^9 K$ the critical angular velocity of this mode is about 
$25\%$ of $(\pi G \bar{\rho})^{1/2}$.


\section{Discussion and Conclusions}

In this paper, we have investigated the properties of 
inertial mode and $r$-mode oscillations 
in rotating isentropic Newtonian stars, using the slow rotation 
approximation.  
By taking account of the effects of the rotational deformation of the 
equilibrium state, we have evaluated the eigenfrequencies of 
the inertial modes and the $r$-modes with an accuracy up to the order of 
$\bar{\Omega}^3$. 
We have also estimated the dissipation timescales due to the 
gravitational radiation and the viscosity 
for a simple neutron star model. 
We show that the inertial modes emit gravitational radiation  
mainly by the mass multipole rather than the current multipole. 
It is also found that the instability 
due to the gravitational radiation reaction is strong
for the most unstable inertial mode, although 
the instability associated with the inertial modes is not as strong as that 
with the $r$-modes.

In spite of the recent improvements in our understanding about the instability 
of the inertial modes and the $r$-modes, it seems that the fundamental 
properties of these modes have not yet been sufficiently
understood, considering that most of the previous investigations of the
inertial modes and the $r$-modes, including the present paper, are restricted 
to
the case of uniformly and slowly rotating, isentropic, Newtonian stars.
In this sense, we do not have any clear understanding about the inertial modes 
and 
the $r$-modes of, for example,
differentially and rapidly rotating, non-isentropic, relativistic stars.
As a first step to relax these restrictions, several studies have already 
appeared.
For example, Andersson, Lockitch \& Friedman (1999) suggest a possibility 
that the assumption of the purely axial mode oscillation leads to inconsistent 
radial behavior for the case of isentropic $relativistic$ stars.
Kojima (1998) and Kojima \& Hosonuma (1999) also argue that the $r$-mode 
spectrum is continuous for relativistic stars.
Therefore it is highly desirable to investigate the properties of the inertial 
modes and
the $r$-modes in the more general case than that investigated previously and
 in this paper,
to conclude that the instability does have an importance in astrophysics.


\acknowledgements

We would like to thank Prof. H. Saio for useful comments.    
S.Y. would like to thank Prof. Y. Eriguchi, Prof. T. Futamase 
and Dr. S. Yoshida for discussions. S.Y. was supported by 
Research Fellowship of the Japan Society for the Promotion of Science for 
Young Scientists.


\appendix


\section{Basic equations for nonradial oscillations 
         of slowly rotating stars}

The derivation of the governing equations of nonradial oscillations of
rotating stars is almost the same as that given by Lee \& Saio (1986) and
Lee (1993), except that the governing equations given in this Appendix 
are formulated to calculate 
the eigenfrequencies and eigenfunctions correct up to the order of $\Omega^3
$.

We introduce column vectors $\bf{y}_1$, $\bf{y}_2$, $\bf{y}_3$, 
$\bf{y}_4$, $\bf{h}$, and $\bf{t}$, whose components are defined by 
\be
y_{1, k} = S_l (a) \, , 
\ee
\be
y_{2, k} = \frac{1}{g a} \,  \delta U_l(a) \, 
\equiv {1\over ga}\left({\delta p_l(a)\over\rho} 
+\delta \Phi_l(a)\right) \, ,
\ee
\be
y_{3, k} = \frac{1}{g a} \,  \delta \Phi_l(a) \, ,
\ee
\be
y_{4, k} = \frac{1}{g} \,  \frac{d \delta \Phi_l(a)}{d a} \, ,
\ee
\be
h_{,k} = H_l (a) \, ,
\ee
and
\be
t_{,k} = T_{l'} (a) \, ,
\ee
where 
\be
g = - \frac{1}{\rho} \, \frac{d p}{d a} (a) \, .
\ee
Here $l=\vert m \vert + 2 k -2 $ and $l'=l+1$ 
for ``even'' modes, and 
$l=\vert m \vert + 2 k -1 $ and $l'=l-1$
for ``odd'' modes, and in both the cases $k$ denote the indices of 
the vectors, and $k = 1, 2, 3, \dots$.

In vector notation, equations of the adiabatic nonradial pulsation
for a slowly rotating star are written as follows:

\noindent
The perturbed mass conservation law ($\ref{mass_con}$) reduces to 
\begin{eqnarray}
a \, \frac{d {\bf y}_1}{da} + \left( 3-\frac{V}{\Gamma} \right) \, {\bf y}_1
+ \frac{V}{\Gamma} ({\bf y}_2-{\bf y}_3) - {\bf \Lambda}_0 {\bf h}  = 
&-& \left( a \, \frac{d \vartheta (\alpha)}{da} \, {\bf 1} + 
  a \, \frac{d \vartheta (\beta)}{da} \, {\bf {\cal A}}_0 \right) {\bf y}_1 
\nonumber\\
&+& 3 \vartheta (\beta) \, {\bf {\cal B}}_0 \, {\bf h} 
+ 3 m \vartheta (\beta) \, {\bf{\cal Q}}_0 \, {\it i} {\bf t} \, .
\label{differ-1}
\end{eqnarray}
The $a$ component of the perturbed Euler's equation 
($\ref{pert_Euler}$) reduces to 
\begin{eqnarray}
a \, \frac{d{\bf y}_2}{da} - ( c_1 \bar{\omega}^2 &+& a A_a ) \, {\bf y}_1 
- (1 - a A_a -U) {\bf y}_2 - a A_a {\bf y}_3 
+ 2 m c_1 \bar{\omega} \bar{\Omega} {\bf h}  
+ 2 c_1 \bar{\omega} \bar{\Omega} {\bf C}_0 \, {\it i} {\bf t} 
\nonumber\\
&=& c_1 \bar{\omega}^2 \, [ 2 \lbrace 
   \eta(\alpha){\bf 1}+\eta(\beta) {\bf{\cal A}}_0 \rbrace \, {\bf y}_1
 - 3 \beta \, {\bf{\cal B}}_0 \, {\bf h}
- 3 m \beta \, {\bf{\cal Q}}_0 \, {\it i} {\bf t} 
\nonumber\\
&& \hspace{0.8in} 
-m \nu \lbrace (\alpha +\eta(\alpha)) {\bf 1} 
      + (\beta +\eta(\beta)){\bf{\cal A}}_0 \rbrace \, {\bf h}
\nonumber\\
&& \hspace{0.8in}
- \nu \lbrace (\alpha +\eta(\alpha)) {\bf C}_0
      + (\beta +\eta(\beta)){\bf{\cal E}}_0 \rbrace \, {\it i} {\bf t} ] \, 
.
\label{differ-2}
\end{eqnarray}
The perturbed Poisson equation ($\ref{pert_Poisson}$) reduces to 
\be
a \, \frac{d{\bf y}_3}{da} - ( 1 - U ) \, {\bf y}_3 - {\bf y}_4 = 0 \, , 
\label{differ-3}
\ee
and 
\begin{eqnarray}
a \, \frac{d{\bf y}_4}{da} &+& a A_a U \, {\bf y}_1 
- U \, \frac{V}{\Gamma} {\bf y}_2 
- \left( {\bf \Lambda}_0 - U \, \frac{V}{\Gamma} {\bf 1} \right) \, 
 {\bf y}_3  + U \, {\bf y}_4 
\nonumber\\
&=& 2 \lbrace \eta(\alpha) {\bf 1} + \eta(\beta) {\bf{\cal A}}_0 
\rbrace \, a \, \frac{d{\bf y}_4}{da} 
+ \lbrace F(\alpha){\bf 1}+F(\beta) \, {\bf{\cal A}}_0 -
     6 \beta ({\bf{\cal A}}_0+{\bf{\cal B}}_0 ) \rbrace \, {\bf y}_4
\nonumber\\
&-&  2 \lbrace \alpha{\bf 1} + \beta {\bf{\cal A}}_0 \rbrace 
\, {\bf \Lambda}_0 {\bf y}_3 \, . \label{differ-4}
\end{eqnarray}
Here, 
\be
U = \frac{d\ln M(a)}{d\ln a} \, , \hspace{0.5in}
V = - \frac{d\ln p}{d\ln a} \, ,
\ee
\be
\vartheta(\alpha) = 3 \alpha + a \frac{d \alpha}{da} \, , \hspace{0.5in}
\eta(\alpha) = \alpha + a \frac{d \alpha}{da} \, ,
\ee
\be
F(\alpha) = 2 U \eta(\alpha) - a \frac{d \alpha}{da} 
+ a \frac{d} {da} \left( a \frac{d \alpha}{da} \right) \, ,
\ee
$\bar{\omega} \equiv \omega/(GM/R^3)^{1/2}$ is the frequency in the unit of 
the Kepler frequency, and $\nu \equiv 2 \Omega / \omega$ . 

\noindent
The $\theta$ and $\varphi$ components of the perturbed Euler's equation 
($\ref{pert_Euler}$) reduce to 
\be 
{\bf\tilde{L}}_0 \, {\bf h} - {\bf\tilde{M}}_1 \, {\it i} {\bf t}
 = \frac{1}{c_1 \bar{\omega}^2} \, {\bf y}_2 + \lbrace 
{\bf\tilde{O}}+{\bf\tilde{M}}_1{\bf\tilde{L}}_1^{-1}{\bf\tilde{K}}
 \rbrace \, {\bf y}_1 \, , \label{alge-1}
\ee
\be 
{\bf\tilde{L}}_1 \, {\it i} {\bf t} - {\bf\tilde{M}}_0 \, {\bf h}
 = - {\bf\tilde{K}} \, {\bf y}_1 \, , \label{alge-2}
\ee
where
\be
{\bf\tilde{L}}_0 = (1+2 \alpha) {\bf L}_0 
 + \beta {\bf \Lambda}_0^{-1} {\bf{\cal H}}_0 - 2 m \nu \beta 
   {\bf \Lambda}_0^{-1} \left( 6 \, {\bf\cal{A}}_0 + {\bf 1} \right) \, ,
\ee
\be
{\bf\tilde{L}}_1 = (1+2 \alpha) {\bf L}_1
 + \beta {\bf \Lambda}_1^{-1} {\bf{\cal  H}}_1 - 2 m \nu \beta
   {\bf \Lambda}_1^{-1} \left( 6 \, {\bf\cal{A}}_1 + {\bf 1} \right) \, ,
\ee
\be
{\bf\tilde{M}}_0 = (1+2 \alpha-2 \beta) {\bf M}_0
 - 6 m \beta {\bf \Lambda}_1^{-1} {\bf{\cal Q}}_1 + 4 \nu \beta
   {\bf \Lambda}_1^{-1} \left( {\bf{\cal D}}_1 {\bf \Lambda}_0 
   + 3 {\bf{\cal E}}_1 + {\bf C}_1 \right) \, ,
\ee
\be
{\bf\tilde{M}}_1 = (1+2 \alpha-2 \beta) {\bf M}_1
 - 6 m \beta {\bf \Lambda}_0^{-1} {\bf{\cal Q}}_0 + 4 \nu \beta
   {\bf \Lambda}_0^{-1} \left( {\bf{\cal D}}_0 {\bf \Lambda}_1 
   + 3 {\bf{\cal E}}_0 + {\bf C}_0 \right) \, ,
\ee
\be
{\bf\tilde{K}} = (1+ \alpha+\eta(\alpha)) \, \nu {\bf K}
 - 3 m \beta {\bf \Lambda}_1^{-1} {\bf{\cal Q}}_1 
 + \nu (\beta+\eta(\beta)) \, {\bf \Lambda}_1^{-1} \, \left( 
  4 {\bf{\cal D}}_1 + {\bf{\cal E}}_1 -2 {\bf{\cal Q}}_1 \right) \, ,
\ee
and
\be
{\bf\tilde{O}} = {\bf \Lambda}_1^{-1} \, \biggl[ m \nu \Bigl\lbrace 
  (1+ \alpha+\eta(\alpha)) \, {\bf 1}
   + (\beta+\eta(\beta)) \, {\bf\cal{A}}_0 \Bigr\rbrace 
 - 3 \beta \left(2 {\bf{\cal A}}_0 + {\bf{\cal B}}_0 \right) \biggr] 
 - {\bf\tilde{M}}_1{\bf\tilde{L}}_1^{-1}{\bf\tilde{K}} \, ;
\ee
\be
{\bf{\cal H}}_0 = 2 {\bf{\cal A}}_0 {\bf \Lambda}_0 
 + 6 {\bf{\cal B}}_0  \, ,
\hspace{0.5in} 
{\bf{\cal H}}_1 = 2 {\bf{\cal A}}_1 {\bf \Lambda}_1
 + 6 {\bf{\cal B}}_1  \, .
\ee

The quantities ${\bf{\cal Q}}_0$, ${\bf{\cal Q}}_1$, ${\bf C}_0$, ${\bf C}_1
$, 
${\bf K}$, ${\bf L}_0$, ${\bf L}_1$, ${\bf \Lambda}_0$, ${\bf \Lambda}_1$, 
${\bf M}_0$, ${\bf M}_1$, ${\bf{\cal A}}_0$, ${\bf{\cal A}}_1$, 
${\bf{\cal B}}_0$, ${\bf{\cal B}}_1$, ${\bf{\cal D}}_0$, 
${\bf{\cal D}}_1$, ${\bf{\cal E}}_0$, and ${\bf{\cal E}}_1$ are matrices 
written as follows:  

\noindent
For even modes, 
\[
({\bf{\cal Q}}_0)_{i,i} = J^m_{l+1} \, , \hspace{.3in} 
({\bf{\cal Q}}_0)_{i+1,i} = J^m_{l+2} \, ,  
\]
\[
({\bf{\cal Q}}_1)_{i,i} = J^m_{l+1} \, , \hspace{.3in}
({\bf{\cal Q}}_1)_{i,i+1} = J^m_{l+2} \, , 
\]
\[
({\bf C}_0)_{i,i} = - (l+2) J^m_{l+1} \, , \hspace{.3in}
({\bf C}_0)_{i+1,i} = (l+1) J^m_{l+2} \, ,
\]
\[
({\bf C}_1)_{i,i} = l J^m_{l+1} \, , \hspace{.3in}
({\bf C}_1)_{i,i+1} = - (l+3) J^m_{l+2} \, ,
\]
\[
({\bf K})_{i,i} = \frac{J^m_{l+1}}{l+1} \, , \hspace{.3in}
({\bf K})_{i,i+1} = - \frac{J^m_{l+2}}{l+2} \, ,
\]
\[
({\bf L}_0)_{i,i} = 1 - \frac{m \nu}{l(l+1)} \, , \hspace{.3in}
({\bf L}_1)_{i,i} = 1 - \frac{m \nu}{(l+1)(l+2)} \, ,
\]
\[
({\bf \Lambda}_0)_{i,i} = l(l+1) \, , \hspace{.3in}
({\bf \Lambda}_1)_{i,i} = (l+1)(l+2) \, ,
\]
\[
({\bf M}_0)_{i,i} = \nu \frac{l}{l+1} \, J^m_{l+1} \, , \hspace{.3in}
({\bf M}_0)_{i,i+1} = \nu \frac{l+3}{l+2} \, J^m_{l+2} \, ,
\]
\[
({\bf M}_1)_{i,i} = \nu \frac{l+2}{l+1} \, J^m_{l+1} \, , \hspace{.3in}
({\bf M}_1)_{i+1,i} = \nu \frac{l+1}{l+2} \, J^m_{l+2} \, ,
\]
\[
({\bf{\cal A}}_0)_{i,i} = - \frac{1}{2} + \frac{3}{2} \, 
\{ (J^m_l)^2 + (J^m_{l+1})^2 \}  \, , \hspace{.3in}
({\bf{\cal A}}_0)_{i+1,i} = ({\bf{\cal A}}_0)_{i,i+1} = 
\frac{3}{2} \, J^m_{l+1} J^m_{l+2} \, ,
\]
\[
({\bf{\cal A}}_1)_{i,i} = - \frac{1}{2} + \frac{3}{2} \, 
\{ (J^m_{l+2})^2 + (J^m_{l+1})^2 \}  \, , \hspace{.3in}
({\bf{\cal A}}_1)_{i+1,i} = ({\bf{\cal A}}_1)_{i,i+1} = 
\frac{3}{2} \, J^m_{l+2} J^m_{l+3} \, ,
\]
\be
({\bf{\cal B}}_0)_{i,i} = l (J^m_{l+1})^2 -(l+1) (J^m_{l})^2 \, , 
\hspace{.3in}
({\bf{\cal B}}_0)_{i+1,i} = l J^m_{l+1} J^m_{l+2} \, ,
\ee
\[
({\bf{\cal B}}_0)_{i,i+1} = - (l+3) J^m_{l+1} J^m_{l+2}  \, , \hspace{.3in}
({\bf{\cal B}}_1)_{i,i} = (l+1) (J^m_{l+2})^2 - (l+2) (J^m_{l+1})^2  \, ,
\]
\[
({\bf{\cal B}}_1)_{i+1,i} = (l+1) J^m_{l+2} J^m_{l+3} \, , \hspace{.3in}
({\bf{\cal B}}_1)_{i,i+1} = - (l+4) J^m_{l+2} J^m_{l+3}  \, ,
\]
\[
({\bf{\cal D}}_0)_{i,i} = J^m_{l+1} \left[ - \frac{1}{2} + \frac{3}{2} \, 
\{ (J^m_l)^2 + (J^m_{l+1})^2  + (J^m_{l+2})^2 \}
\right]  \, , \hspace{.1in}
({\bf{\cal D}}_0)_{i,i+1} = \frac{3}{2} \, J^m_{l+1} J^m_{l+2} J^m_{l+3}  \,
 ,
\]
\[
({\bf{\cal D}}_0)_{i+1,i} = J^m_{l+2} \left[ - \frac{1}{2} + \frac{3}{2} \, 
\{ (J^m_{l+1})^2 + (J^m_{l+2})^2  + (J^m_{l+3})^2 \}
\right]  \, , \hspace{.1in}
({\bf{\cal D}}_0)_{i+2,i} = \frac{3}{2} \, J^m_{l+2} J^m_{l+3} J^m_{l+4}  \,
 ,
\]
\[
({\bf{\cal D}}_1)_{i,i} = J^m_{l+1} \left[ - \frac{1}{2} + \frac{3}{2} \, 
\{ (J^m_l)^2 + (J^m_{l+1})^2  + (J^m_{l+2})^2 \}
\right]  \, , \hspace{.1in}
({\bf{\cal D}}_1)_{i+1,i} = \frac{3}{2} \, J^m_{l+1} J^m_{l+2} J^m_{l+3}  \,
 ,
\]
\[
({\bf{\cal D}}_1)_{i,i+1} = J^m_{l+2} \left[ - \frac{1}{2} + \frac{3}{2} \, 
\{ (J^m_{l+1})^2 + (J^m_{l+2})^2  + (J^m_{l+3})^2 \}
\right]  \, , \hspace{.1in}
({\bf{\cal D}}_1)_{i,i+2} = \frac{3}{2} \, J^m_{l+2} J^m_{l+3} J^m_{l+4}  \,
 ,
\]
\[
({\bf{\cal E}}_0)_{i,i} = J^m_{l+1} \left[ \frac{1}{2} \, (l+2) 
+ \frac{3}{2} \, 
\{ (l+1) (J^m_{l+2})^2 - (l+2) ( (J^m_l)^2 + (J^m_{l+1})^2 ) \} \right] \, ,
\]
\[
({\bf{\cal E}}_0)_{i+1,i} = J^m_{l+2}\left[ -\frac{1}{2} \, (l+1) 
 +\frac{3}{2} \, 
\{ (l+1) ( (J^m_{l+2})^2 + (J^m_{l+3})^2 ) - (l+2) (J^m_{l+1})^2\} \right] 
\, ,
\]
\[
({\bf{\cal E}}_0)_{i,i+1} = -\frac{3}{2} \, (l+4) J^m_{l+1} J^m_{l+2} J^m_{l
+3} 
\, , \hspace{.3in}
({\bf{\cal E}}_0)_{i+2,i} = \frac{3}{2} \, (l+1) J^m_{l+2} J^m_{l+3} J^m_{l+
4}  
\, ,
\]
\[
({\bf{\cal E}}_1)_{i,i} = J^m_{l+1} \left[- \frac{1}{2} \, l 
+ \frac{3}{2} \,
\{ l ((J^m_{l+1})^2+ (J^m_{l+2})^2) - (l+1) (J^m_{l})^2 \} \right] \, ,
\]
\[
({\bf{\cal E}}_1)_{i,i+1} = J^m_{l+2} \left[\frac{1}{2} (l+3) + \frac{3}{2} 
\,
\{ (l+2)  (J^m_{l+3})^2 - (l+3)( (J^m_{l+1})^2+(J^m_{l+2})^2) \} \right] \, 
,
\]
\[
({\bf{\cal E}}_1)_{i+1,i} = \frac{3}{2} \, l J^m_{l+1} J^m_{l+2} J^m_{l+
3}
\, , \hspace{.3in}
({\bf{\cal E}}_1)_{i,i+2} = -\frac{3}{2} \, (l+5) J^m_{l+2} J^m_{l+3} J^m_{l
+4}
\, ,
\]
where $l=\vert m \vert + 2 i -2 $, $i = 1,2,3,\dots $, and
\be
J^m_l \equiv \left[ \frac{(l+m)(l-m)}{(2l-1)(2l+1)} \right]^{\half} 
\, .  \label{J_l}
\ee

\noindent
For odd modes, 
\[
({\bf{\cal Q}}_0)_{i,i} = J^m_{l} \, , \hspace{.3in}
({\bf{\cal Q}}_0)_{i,i+1} = J^m_{l+1} \, ,
\]
\[
({\bf{\cal Q}}_1)_{i,i} = J^m_{l} \, , \hspace{.3in}
({\bf{\cal Q}}_1)_{i+1,i} = J^m_{l+1} \, ,
\]
\[
({\bf C}_0)_{i,i} = (l-1) J^m_{l} \, , \hspace{.3in}
({\bf C}_0)_{i+1,i} = -(l+2) J^m_{l+1} \, ,
\]
\[
({\bf C}_1)_{i,i} = -(l+1) J^m_{l} \, , \hspace{.3in}
({\bf C}_1)_{i+1,i} = l J^m_{l+1} \, ,
\]
\[
({\bf K})_{i,i} = - \frac{J^m_l}{l} \, , \hspace{.3in}
({\bf K})_{i+1,i} = \frac{J^m_{l+1}}{l+1} \, ,
\]
\[
({\bf L}_0)_{i,i} = 1 - \frac{m \nu}{l(l+1)} \, , \hspace{.3in}
({\bf L}_1)_{i,i} = 1 - \frac{m \nu}{l(l-1)} \, ,
\]
\[
({\bf \Lambda}_0)_{i,i} = l(l+1) \, , \hspace{.3in}
({\bf \Lambda}_1)_{i,i} = l(l-1) \, ,
\]
\[
({\bf M}_0)_{i,i} = \nu \frac{l+1}{l} \, J^m_l \, , \hspace{.3in}
({\bf M}_0)_{i+1,i} = \nu \frac{l}{l+1} \, J^m_{l+1} \, ,
\]
\[
({\bf M}_1)_{i,i} = \nu \frac{l-1}{l} \, J^m_l \, , \hspace{.3in}
({\bf M}_1)_{i,i+1} = \nu \frac{l+2}{l+1} \, J^m_{l+1} \, ,
\]
\[
({\bf{\cal A}}_0)_{i,i} = - \frac{1}{2} + \frac{3}{2} \,
\{ (J^m_{l+1})^2 + (J^m_{l})^2 \}  \, , \hspace{.3in}
({\bf{\cal A}}_0)_{i+1,i} = ({\bf{\cal A}}_0)_{i,i+1} =
\frac{3}{2} \, J^m_{l+1} J^m_{l+2} \, ,
\]
\[
({\bf{\cal A}}_1)_{i,i} = - \frac{1}{2} + \frac{3}{2} \,
\{ (J^m_{l-1})^2 + (J^m_l)^2 \}  \, , \hspace{.3in}
({\bf{\cal A}}_1)_{i+1,i} = ({\bf{\cal A}}_1)_{i,i+1} =
\frac{3}{2} \, J^m_{l} J^m_{l+1} \, ,
\]
\be
({\bf{\cal B}}_0)_{i,i} = l (J^m_{l+1})^2 -(l+1) (J^m_{l})^2 \, ,
\hspace{.3in}
({\bf{\cal B}}_0)_{i+1,i} = l J^m_{l+1} J^m_{l+2} \, ,
\ee
\[
({\bf{\cal B}}_0)_{i,i+1} = - (l+3) J^m_{l+1} J^m_{l+2}  \, , \hspace{.3in}
({\bf{\cal B}}_1)_{i,i} = (l-1) (J^m_l)^2 - l (J^m_{l-1})^2  \, ,
\]
\[
({\bf{\cal B}}_1)_{i+1,i} = (l-1) J^m_l J^m_{l+1} \, , \hspace{.3in}
({\bf{\cal B}}_1)_{i,i+1} = - (l+2) J^m_l J^m_{l+1}  \, ,
\]
\[
({\bf{\cal D}}_0)_{i,i} = J^m_l \left[ - \frac{1}{2} + \frac{3}{2} \,
\{ (J^m_{l-1})^2 + (J^m_l)^2  + (J^m_{l+1})^2 \}
\right]  \, , \hspace{.1in}
({\bf{\cal D}}_0)_{i+1,i} = \frac{3}{2} \, J^m_l J^m_{l+1} J^m_{l+2}  \, ,
\]
\[
({\bf{\cal D}}_0)_{i,i+1} = J^m_{l+1} \left[ - \frac{1}{2} + \frac{3}{2} \,
\{ (J^m_l)^2 + (J^m_{l+1})^2  + (J^m_{l+2})^2 \}
\right]  \, , \hspace{.1in}
({\bf{\cal D}}_0)_{i,i+2} = \frac{3}{2} \, J^m_{l+1} J^m_{l+2} J^m_{l+3}  \,
 ,
\]
\[
({\bf{\cal D}}_1)_{i,i} = J^m_l \left[ - \frac{1}{2} + \frac{3}{2} \,
\{ (J^m_{l-1})^2 + (J^m_l)^2  + (J^m_{l+1})^2 \}
\right]  \, , \hspace{.1in}
({\bf{\cal D}}_1)_{i,i+1} = \frac{3}{2} \, J^m_l J^m_{l+1} J^m_{l+2}  \, ,
\]
\[
({\bf{\cal D}}_1)_{i+1,i} = J^m_{l+1} \left[ - \frac{1}{2} + \frac{3}{2} \,
\{ (J^m_l)^2 + (J^m_{l+1})^2  + (J^m_{l+2})^2 \}
\right]  \, , \hspace{.1in}
({\bf{\cal D}}_1)_{i+2,i} = \frac{3}{2} \, J^m_{l+1} J^m_{l+2} J^m_{l+3}  \,
 ,
\]
\[
({\bf{\cal E}}_0)_{i,i} = J^m_l \left[ -\frac{1}{2} (l-1) + \frac{3}{2} \,
\{ (l-1) ((J^m_l)^2 + (J^m_{l+1})^2) - l (J^m_{l-1})^2 \} \right] \, ,
\]
\[
({\bf{\cal E}}_0)_{i,i+1} = J^m_{l+1} \left[\frac{1}{2} (l+2) + \frac{3}{2} 
\,
\{ -(l+2) ( (J^m_l)^2 + (J^m_{l+1})^2 ) + (l+1) (J^m_{l+2})^2\} \right] \, ,
\]
\[
({\bf{\cal E}}_0)_{i+1,i} = \frac{3}{2} \, (l-1) J^m_l J^m_{l+1} J^m_{l+2}
\, , \hspace{.3in}
({\bf{\cal E}}_0)_{i,i+2} = -\frac{3}{2} \, (l+4) J^m_{l+1} J^m_{l+2} J^m_{l
+3}
\, ,
\]
\[
({\bf{\cal E}}_1)_{i,i} = J^m_l \left[ \frac{1}{2} \, (l+1) 
+ \frac{3}{2} \,
\{ l (J^m_{l+1})^2 - (l+1) ((J^m_{l-1})^2 + (J^m_l)^2)\} \right] \, ,
\]
\[
({\bf{\cal E}}_1)_{i+1,i} = J^m_{l+1} \left[-\frac{1}{2} l + \frac{3}{2} \,
\{ l ((J^m_{l+1})^2 +(J^m_{l+2})^2)- (l+1) (J^m_l)^2 \} \right] \, ,
\]
\[
({\bf{\cal E}}_1)_{i,i+1} = -\frac{3}{2} \, (l+3) J^m_l J^m_{l+1} J^m_{l+2}
\, , \hspace{.3in}
({\bf{\cal E}}_1)_{i+2,i} = \frac{3}{2} \, l J^m_{l+1} J^m_{l+2} J^m_{l+
3}
\, ,
\]
where $l=\vert m \vert + 2 i -1 $, $i = 1,2,3,\dots $.

Eliminating $\bf h$ and 
$\it i \bf t$ from equations (\ref{differ-1}) and (\ref{differ-2}) by 
using equations (\ref{alge-1}) and (\ref{alge-2}), 
equations (\ref{differ-1})--(\ref{differ-4}) reduce to a set of first-order 
linear ordinary  differential equations for ${\bf y}_1$, ${\bf y}_2$, 
${\bf y}_3$ and ${\bf y}_4$ as follows:
\be
a\, \frac{d {\bf y}_1}{da} = 
 \left\lbrace \left( \frac{V}{\Gamma} -3 \right) \, {\bf 1} 
 + {\bf{\cal F}}_{11} \right\rbrace \, {\bf y}_1 + \left( 
\frac{{\bf{\cal F}}_{12}}{c_1 \bar{\omega}^2} - \frac{V}{\Gamma}{\bf 1} 
\right) 
\, {\bf y}_2 + \frac{V}{\Gamma} \, {\bf y}_3 \, , \label{basic-eq1}
\ee
\be
a\, \frac{d {\bf y}_2}{da} = \Bigl\lbrace (c_1 \bar{\omega}^2 + a A_a) 
 \, {\bf 1} + c_1 \bar{\omega}^2 {\bf{\cal F}}_{21} \Bigr\rbrace \, {\bf y}_
1 
+ \Bigl\lbrace ( 1 - a A_a - U ){\bf 1} + {\bf{\cal F}}_{22} \Bigr\rbrace 
 \, {\bf y}_2 + a A_a \, {\bf y}_3 \, , \label{basic-eq2}
\ee
\be
a\, \frac{d {\bf y}_3}{da} = ( 1 - U ) \, {\bf y}_3  + {\bf y}_4 \, , 
\label{basic-eq3}
\ee
\begin{eqnarray}
a\, \frac{d {\bf y}_4}{da} = -a A_a U {\bf{\cal J}}^{-1} {\bf y}_1 
&+& U \, \frac{V}{\Gamma} {\bf{\cal J}}^{-1} {\bf y}_2 
 + {\bf{\cal J}}^{-1} 
   \left\lbrace {\bf \Lambda}_0 -  U \, \frac{V}{\Gamma} {\bf 1} 
   - 2 ( \alpha {\bf 1} + \beta {\bf{\cal A}}_0 ) {\bf \Lambda}_0 
 \right\rbrace \, {\bf y}_3  \nonumber \\ 
&+&  {\bf{\cal J}}^{-1} 
 \Bigl\lbrace -U {\bf 1} + F(\alpha) {\bf 1} + F(\beta) {\bf{\cal A}}_0 
   - 6 \beta ( {\bf{\cal A}}_0 + {\bf{\cal B}}_0 )  \Bigr\rbrace 
 \, {\bf y}_4 \, , \label{basic-eq4}
\end{eqnarray}
where 
\be
{\bf{\cal F}}_{11} = \Bigl\lbrace {\bf 1} + 3 \vartheta (\beta) {\bf{\cal P}
} 
\Bigr\rbrace \, \tilde{\bf W} \tilde{\bf O} 
 - a \, \frac{d \vartheta (\alpha)}{da} {\bf 1} 
 - a \, \frac{d \vartheta (\beta)}{da} {\bf{\cal A}}_0 
 - 3m \vartheta (\beta) {\bf{\cal Q}}_0 \tilde{{\bf L}}_1^{-1} 
   \tilde{{\bf K}} \, ,
\ee
\be
{\bf{\cal F}}_{12} = \Bigl\lbrace {\bf 1} + 3 \vartheta (\beta) {\bf{\cal P}
}
\Bigr\rbrace \, \tilde{\bf W} \, ,
\ee
\begin{eqnarray}
{\bf{\cal F}}_{21} &=& 2 \Bigl\lbrace \eta (\alpha) {\bf 1} 
   + \eta (\beta) {\bf{\cal A}}_0 \Bigr\rbrace 
 - {\bf{\cal R}} \tilde{\bf W} \tilde{\bf O} 
\nonumber \\
&& +  \biggl[ \nu \Bigl\lbrace (1+\alpha+\eta(\alpha)) {\bf C}_0 
   + (\beta+\eta(\beta)) {\bf{\cal E}}_0 \Bigr\rbrace 
   + 3m \beta {\bf{\cal Q}}_0 \Biggr] 
     \tilde{{\bf L}}_1^{-1} \tilde{{\bf K}} \, ,
\end{eqnarray}
\be
{\bf{\cal F}}_{22} = - {\bf{\cal R}} \, \tilde{\bf W} \, ,
\ee
\be
{\bf{\cal P}} = \Bigl( {\bf{\cal B}}_0  
 + m {\bf{\cal Q}}_0 \tilde{{\bf L}}_1^{-1} \tilde{{\bf M}}_0 \Bigr) 
{\bf \Lambda}_0^{-1} \, ,
\ee
\begin{eqnarray}
{\bf{\cal R}} &=& \biggl[ m \nu \Bigl\lbrace 
 (1+\alpha+\eta(\alpha)){\bf 1}+(\beta+\eta(\beta)) {\bf{\cal A}}_0  
 \Bigr\rbrace + 3 \beta {\bf{\cal B}}_0 \biggr] {\bf \Lambda}_0^{-1} 
\nonumber \\
&& + \biggl[ \nu \Bigl\lbrace (1+\alpha+\eta(\alpha)){\bf C}_0 
  + (\beta+\eta(\beta)) {\bf{\cal E}}_0 + 3m \beta {\bf{\cal Q}}_0 
 \biggr] \tilde{{\bf L}}_1^{-1} \tilde{{\bf M}}_0 
{\bf \Lambda}_0^{-1} \, ,
\end{eqnarray}
\be
\tilde{\bf W} = {\bf \Lambda}_0 (\tilde{\bf L}_0 
 - \tilde{\bf M}_1 \tilde{\bf L}_1^{-1} \tilde{\bf M}_0 )^{-1} \, ,
\ee
\be
{\bf{\cal J}} = {\bf 1} - 2 \lbrace \eta (\alpha) {\bf 1} 
 + \eta (\beta) {\bf{\cal A}}_0 \rbrace \, .
\ee

The surface boundary conditions are 
\be
{\bf y}_1 - {\bf y}_2 + {\bf y}_3 = 0 \, , \label{boundary-1}
\ee
and 
\be
\biggl[ {\bf \Lambda}_2 \Bigl\{ (1+\alpha){\bf 1} + \beta {\bf{\cal A}}_0 
 \Bigr\} +\Bigl\{\alpha {\bf 1} + \beta {\bf{\cal A}}_0 \Bigr\} 
 {\bf \Lambda}_0 \biggr] {\bf y}_3 + 
\Bigl\{ {\bf \Lambda}_2 (\alpha {\bf 1} + \beta {\bf{\cal A}}_0) 
+ {\bf 1} \Bigr\} {\bf y}_4  = 0 \, , \label{boundary-2}
\ee
where ${\bf \Lambda}_2$ is a matrix whose elements are given by 
\be
({\bf \Lambda}_2)_{i,i} = (l+1) \, .
\ee
Here  $l=\vert m \vert + 2i-2$ for ``even'' modes, 
$l=\vert m \vert + 2i-1$ for ``odd'' modes. 
The inner boundary conditions at the stellar center are the regularity 
conditions of the eigenfunctions.


%


\newpage

\figcaption{The first two expansion coefficients $S_l(a)$ [panel a], 
$H_l(a)$ [panel b], ${\it i} T_l(a)$ [panel c] for the even parity 
inertial modes with $m=2$ and $l_0 - \vert m \vert = 2 $
 are plotted against $a/R$ for a polytropic model 
with the index $n=1$, where we assume $\bar{\Omega}=0.01$. 
The solid curves give the expansion coefficients corresponding to the 
mode with $\kappa_0=1.100$, while the dotted curves correspond to the mode 
having $\kappa_0=-0.557$. 
The eigenfunctions are normalized so that ${\it i}T_3(a=R)=1$. 
Attached labels $l=2,3,4,5,\cdots$ denote the spherical harmonic indices 
associated with the expansion coefficients.
\label{ef_p1}}

\figcaption{The first three expansion coefficients  $S_l(a)$ [panel a], 
$H_l(a)$ [panel b], ${\it i} T_l(a)$ [panel c] for the $m=2$ even parity 
inertial mode with $\kappa_0=1.520$ and $l_0 - \vert m \vert = 4 $ 
are plotted against $a/R$ for a polytropic model with 
the index $n=1$, where $\bar{\Omega}=0.01$.  The eigenfunctions are 
normalized so that ${\it i}T_3(a=R)=1$. Attached labels $l=2,3,4,5,\cdots$ 
denote the spherical harmonic indices associated with the expansion 
coefficients.  
\label{ef_a1}}

\figcaption{The same as Figure 2, but for the mode having eigenvalue 
$\kappa_0=0.863$. 
\label{ef_a2}}

\figcaption{Critical angular velocities $\Omega_c$
for the $r$-mode and two inertial modes 
are plotted against the temperature of a neutron star. The curve  
labeled  ``$r$-mode'' is for the $m=2$ $r$-mode with $\kappa_0=0.667$. 
The curve labeled ``even'' is for the $m=2$ even parity inertial mode with 
$\kappa_0=1.100$.  
The curve labeled ``odd'' is for 
the $m=2$ odd parity inertial mode with $\kappa_0=0.517$. 
Here, the critical angular velocity is normalized by using 
$(\pi G \bar{\rho})^{1/2}$ and the unit of the temperature is Kelvin. \label
{fig5}}


\clearpage

\begin{deluxetable}{crrr}
\footnotesize
\tablecaption{Eigenvalues $(\kappa_0, \kappa_2)\tablenotemark{a}$ of inertial 
modes 
and $r$-modes for the $p=K\rho^2$ Polytrope. \label{n1pol}}
\tablewidth{0pt}
\tablehead{
\colhead{$l_0-\vert m \vert$\tablenotemark{b}}   &
\colhead{$m=1$}   & \colhead{$m=2$}  & \colhead{$m=3$}
}
\startdata
 1\tablenotemark{c}
  & ( 1.00000,\ -0.00018)\phm{*} & ( 0.66667,\  0.39827)*
  & ( 0.50000,\  0.42727)*       \nl
 2& (-0.40137,\  0.20020)\phm{*} & (-0.55659,\  0.12248)\phm{*}
  & (-0.63164,\  0.05993)\phm{*} \nl
  & ( 1.41300,\  0.23767)\phm{*} & ( 1.10003,\  0.46560)*
  & ( 0.90491,\  0.54494)*       \nl
 3& (-1.03238,\ -0.21090)\phm{*} & (-1.02588,\ -0.22877)\phm{*}
  & (-1.01487,\ -0.23750)\phm{*} \nl
  & ( 0.69059,\  0.36221)*       & ( 0.51734,\  0.37950)*
  & ( 0.41265,\  0.35766)*       \nl
  & ( 1.61373,\  0.26999)\phm{*} & ( 1.35778,\  0.45492)*
  & ( 1.17674,\  0.55092)*       \nl
 4& (-1.31227,\ -0.32577)\phm{*} & (-1.27289,\ -0.33970)\phm{*}
  & (-1.23863,\ -0.34435)\phm{*} \nl
  & (-0.17879,\  0.01402)\phm{*} & (-0.27533,\  0.00051)\phm{*}
  & (-0.33327,\ -0.01537)\phm{*} \nl
  & ( 1.05153,\  0.45459)\phm{*} & ( 0.86295,\  0.50140)*
  & ( 0.73430,\  0.50562)*       \nl
  & ( 1.72626,\  0.24391)\phm{*} & ( 1.51957,\  0.41210)*
  & ( 1.36056,\  0.51593)*       \nl

\enddata

\tablenotetext{a}{$\Omega(\kappa_0+\kappa_2\bar{\Omega}^2)=\omega$ is 
the mode frequency in the corotating frame up to the order of 
$\bar{\Omega}^3$.
The modes marked with an asterisk $\ast$ satisfy the condition
$\sigma(\sigma+m\Omega)<0$ to the lowest order in $\Omega$, and are 
considered to be unstable to the gravitational radiation reaction in 
the absence of viscous dissipation.}
\tablenotetext{b}{Our definition of the angular quantum number $l_0$ is 
not the same as that of Lockitch \& Friedman (1998), but the same as that 
of Lindblom \& Ipser (1998). Eigenmodes having  odd (even) values of 
$l_0-\vert m \vert$ are odd (even) parity modes. }
\tablenotetext{c}{These modes are $r$-modes. To the lowest order in 
$\Omega$, $\kappa_0=2/(m+1)$ (Papalouizou \& Prigle 1978) and 
is independent of the equation of state.}
\end{deluxetable}

\clearpage

\begin{deluxetable}{crrrr}
\footnotesize
\tablecaption{Eigenvalues 
$(\kappa_0,\ \kappa_2)\tablenotemark{a}$ \  
of $m=2$ inertial modes and $r$-modes
for several values of the polytropic index $n$. \label{m2pol}}
\tablewidth{0pt}
\tablehead{
\colhead{$l_0-\vert m \vert$\tablenotemark{b}} & 
\colhead{$n=0$}   & \colhead{$n=1$}  & \colhead{$n=1.5$}   & 
\colhead{$n=2$}   
} 
\startdata
 1\tablenotemark{c}
  & ( 0.66667, 0.76543)*       & ( 0.66667, 0.39827)* 
  & ( 0.66667, 0.28248)*       & ( 0.66667, 0.19701)*       \nl
 2& (-0.23193, 0.02930)\phm{*} & (-0.55659, 0.12248)\phm{*} 
  & (-0.69646, 0.13213)\phm{*} & (-0.82771, 0.12773)\phm{*} \nl
  & ( 1.23193, 0.88145)*       & ( 1.10003, 0.46560)* 
  & ( 1.06259, 0.33590)*       & ( 1.03438, 0.24322)*       \nl
 3& (-0.76334,-0.55320)\phm{*} & (-1.02588,-0.22877)\phm{*} 
  & (-1.12777,-0.15318)\phm{*} & (-1.21712,-0.11767)\phm{*} \nl
  & ( 0.46690, 0.62847)*       & ( 0.51734, 0.37950)* 
  & ( 0.53563, 0.29081)*       & ( 0.55163, 0.22228)*       \nl
  & ( 1.49644, 0.77599)*       & ( 1.35778, 0.45492)* 
  & ( 1.31003, 0.34136)*       & ( 1.27054, 0.25749)*       \nl
 4& (-1.09257,-0.77973)\phm{*} & (-1.27289,-0.33970)\phm{*} 
  & (-1.34194,-0.24015)\phm{*} & (-1.40134,-0.18109)\phm{*} \nl
  & (-0.10179, 0.00439)\phm{*} & (-0.27533, 0.00051)\phm{*} 
  & (-0.36424,-0.01189)\phm{*} & (-0.45689,-0.02235)\phm{*} \nl
  & ( 0.88425, 0.89693)*       & ( 0.86295, 0.50140)* 
  & ( 0.85863, 0.38371)*       & ( 0.85688, 0.29910)*       \nl
  & ( 1.64344, 0.63981)*       & ( 1.51957, 0.41210)* 
  & ( 1.47219, 0.32232)*       & ( 1.43071, 0.25292)*       \nl

\enddata
 
\tablenotetext{a}{$\Omega(\kappa_0+\kappa_2\bar{\Omega}^2)=\omega$ is
the mode frequency in the corotating frame up to the order of 
$\bar{\Omega}^3$. The modes marked with an asterisk $\ast$ satisfy the 
condition $\sigma(\sigma+m\Omega)<0$ to the lowest order in $\Omega$, and 
are considered to be unstable to the gravitational radiation reaction 
the absence of viscous dissipation.}
\tablenotetext{b}{Our definition of the angular quantum number $l_0$ is not 
the same as that of Lockitch \& Friedman (1998), but the same as that of 
Lindblom \& Ipser (1998). Eigenmodes having  odd (even) values of 
$l_0-\vert m \vert$ are odd (even) parity modes. }
\tablenotetext{c}{These modes are $r$-modes. To the lowest order in 
$\Omega$, $\kappa_0=2/(m+1)$ (Papalouizou \& Prigle 1978) and 
is independent of the equation of state.}
\end{deluxetable}

\clearpage

\begin{deluxetable}{ccccc}
\footnotesize
\tablecaption{Node number of the dominating expansion coefficients 
of $\delta v^i$  for $r$-modes and inertial modes. \label{node}}
\tablewidth{0pt}
\tablehead{
  \colhead{$l_0-\vert m \vert$\tablenotemark{a}} 
& \colhead{$1$}   & \colhead{$2$}  & \colhead{$3$}   & \colhead{$4$}
}
\startdata
 node number of $S_{\vert m \vert\phm{+1}}\tablenotemark{b}$&$\cdots$& 1      
&$\cdots$      
 & 2        \nl
 node number of $S_{\vert m \vert+1}      $&$\cdots$&$\cdots$& 1
 &$\cdots$  \nl
 node number of $S_{\vert m \vert+2}      $&$\cdots$&$\cdots$&$\cdots$      
 & 1        \nl
 node number of $S_{\vert m \vert+3}      $&$\cdots$&$\cdots$&$\cdots$
 &$\cdots$  \nl
 & & & & \nl
 node number of $H_{\vert m \vert\phm{+1}}$&$\cdots$& 1      &$\cdots$      
 & 2        \nl
 node number of $H_{\vert m \vert+1}      $&$\cdots$&$\cdots$& 1
 &$\cdots$  \nl
 node number of $H_{\vert m \vert+2}      $&$\cdots$&$\cdots$&$\cdots$      
 & 1        \nl
 node number of $H_{\vert m \vert+3}      $&$\cdots$&$\cdots$&$\cdots$
 &$\cdots$  \nl
 & & & & \nl
 node number of ${\it i}T_{\vert m \vert\phm{+1}}$& 0      &$\cdots$& 1      
 
&$\cdots$\nl
 node number of ${\it i}T_{\vert m \vert+1}      $&$\cdots$& 0      &$\cdots
$
 & 1       \nl
 node number of ${\it i}T_{\vert m \vert+2}      $&$\cdots$&$\cdots$& 0      
 
&$\cdots$ \nl
 node number of ${\it i}T_{\vert m \vert+3}      $&$\cdots$&$\cdots$&$\cdots
$
 & 0       \nl

\enddata

\tablenotetext{a}{Our definition of angular quantum number $l_0$ is not the 
same
as that of Lockitch and Friedman (1998), but that of Lindblom and Ipser
(1998). Eigenmodes having  odd (even) values of $l_0-\vert m \vert$ are
corresponding to odd (even) parity modes. }
\tablenotetext{b}{For expansion coefficients $S_l$, we include the node at 
the stellar surface in the count of nodes.}

\end{deluxetable}

\clearpage

\begin{deluxetable}{ccccccccc}
\footnotesize
\tablecaption{Dissipative timescales of inertial modes\tablenotemark{a} \ 
and $r$-modes for $n=1$ polytropic model at $T=10^9K$ and $\Omega=\sqrt{\pi 
G \bar{\rho}}$.
\label{times_m2e}}
\tablewidth{0pt}
\tablehead{
\colhead{$m$} & \colhead{$l_0 -\vert m \vert$} & \colhead{$\kappa_0$}
 & \colhead{$\tilde \tau_B$(s)}     & \colhead{$\tilde \tau_S$(s)}     
 & \colhead{$\tilde \tau_{2}$(s)\tablenotemark{b}} 
 & \colhead{$\tilde \tau_{3}$(s)} 
 & \colhead{$\tilde \tau_{4}$(s)}
 & \colhead{$\tilde \tau_{5}$(s)} 
} 
\startdata
1& 3& 0.691 &$ 5.89\times 10^{9} $&$  9.23\times 10^{7} $
 &$ -2.46\times 10^{5}\da  $&$ -1.27\times 10^{8}\phm{\da} $&$ \cdots$
&$\dots$\nl
 & & & & & & & & \nl
2&1\tablenotemark{c}
 &0.667 &$ 2.03\times 10^{11} $&$ 2.50\times 10^{8} $
 &$ -3.31\times 10^{0}\phm{\da} $&$ -3.49\times 10^{2}\da$&$ \cdots$
&$\dots$\nl
 &2& 1.100 &$ 3.35\times 10^{9} $&$ 1.23\times 10^{8} $
 &$ -1.71\times 10^{3}\da $&$ -3.43\times 10^{4}\phm{\da} $&$ \cdots$
&$\dots$\nl
 &3&0.517 &$ 6.47\times 10^{9}  $&$ 6.18\times 10^{7} $
 &$ -1.31\times 10^{8}\phm{\da} $&$ -8.39\times 10^{4}\da $
&$ -1.88\times 10^{6}\phm{\da} $&$\dots$\nl
 & &1.358 &$ 4.13\times 10^{9}  $&$ 7.14\times 10^{7} $
 &$ -7.12\times 10^{9}\phm{\da} $&$ -1.69\times 10^{7}\da $
&$ -1.63\times 10^{9}\phm{\da} $&$\dots$\nl
 &4& 0.863 &$ 1.94\times 10^{9} $&$ 4.92\times 10^{7} $
 &$ -1.63\times 10^{5}\da $&$ -1.12\times 10^{7}\phm{\da} $&$ \cdots$
&$\dots$\nl
 & & 1.520 &$ 4.82\times 10^{9} $&$ 4.87\times 10^{7} $
 &$ -3.35\times 10^{7}\da $&$ -1.90\times 10^{11}\phm{\da} $&$ \cdots$
&$\dots$\nl
 & & & & & & & & \nl
3&1\tablenotemark{c}
 &0.500 &$ 6.63\times 10^{10} $&$ 1.43\times 10^{8} $
 &$\dots$&$ -3.17\times 10^{1}\phm{\da} $&$ -1.88\times 10^{3}\da $
&$\dots$\nl
 &2& 0.905 &$ 1.88\times 10^{9} $&$ 9.41\times 10^{7}  $
 &$\dots$&$ -8.62\times 10^{3}\da $&$ -2.77\times 10^{4}\phm{\da} $
&$\dots$\nl
 &3&0.413 &$ 6.97\times 10^{9}  $&$ 4.78\times 10^{7} $
 &$\dots$&$ -1.86\times 10^{10}\phm{\da} $&$ -5.30\times 10^{5}\da $
&$ -4.07\times 10^{6}\phm{\da} $\nl
 & & 1.177 &$ 2.49\times 10^{9}  $&$ 6.11\times 10^{7} $
 &$\dots$&$ -1.65\times 10^{11}\phm{\da} $&$ -5.11\times 10^{6}\da $
&$ -6.11\times 10^{7}\phm{\da} $\nl
 &4& 0.734 &$ 1.29\times 10^{9} $&$ 4.10\times 10^{7}  $
 &$\dots$&$ -2.00\times 10^{6}\da $&$ -1.16\times 10^{7}\phm{\da} $
&$\dots$\nl
 & & 1.361 &$ 3.06\times 10^{9} $&$ 4.18\times 10^{7}  $
 &$\dots$&$ -3.45\times 10^{7}\da $&$ -1.75\times 10^{10}\phm{\da} $
&$\dots$\nl
\enddata

\tablenotetext{a}{We present dissipative timescales only for those modes 
that are unstable to gravitational radiation reaction.}
\tablenotetext{b}{We present dissipative timescales $\tilde{\tau}_l$ due to 
the gravitational radiation reaction only for those of  
the dominant multipole moments, 
where 
for modes with odd $l_0$ 
$\tilde{\tau}_{2k+1}=\tilde{\tau}_{D,2k+1}$ and 
$\tilde{\tau}_{2k}=\tilde{\tau}_{J,2k}$, and for modes with even
$l_0$ $\tilde{\tau}_{2k+1}=\tilde{\tau}_{J,2k+1}$ and 
$\tilde{\tau}_{2k}=\tilde{\tau}_{D,2k}$. Here $k=1,2,3,\dots$. 
Dissipative timescales due to the mass multipole radiation are 
marked with a dagger $\dagger$. 
}
\tablenotetext{c}{This mode is the $r$-mode. Since the expansion of
perturbations vanishes to the lowest order in $\Omega$ for $r$-modes, 
the definition of the bulk viscosity timescale $\tilde{\tau}_B$ for 
this mode is not the same as that of equation (\ref{tau2}), but that of 
Lindblom et al. (1999).}
 
\end{deluxetable}


\begin{thebibliography}{}

\bibitem[Andersson 1998]{nils97}
 Andersson, N. 1998, \apj, 502, 708
%
\bibitem[Andersson et al.\ 1998]{aks98}
 Andersson, N., Kokkotas, K. and Schutz B. F. 1998, \apj, 510, 846
%
\bibitem[Andersson et al.\ 1999]{alf99}
 Anderson, N., Locktich, K. H. and Friedman, J. L. 1999, in preparation
%
\bibitem[Berthomiue et al. 1978]{be78}
Berthomieu, G., Gonczi, G., Graff, Ph., Provost, J., and Rocca, A. 1978, 
\aap, 70, 597  
%
\bibitem[Beyer and Kokkotas 1999]{bk99}
 Beyer, H. R. and Kokkotas, K. D. 1999, ``On the r-mode spectrum of 
 relativistic stars'' [gr-qc/9903019]
%
\bibitem[Bryan 1889]{b1889}
 Bryan, G. H. 1889, Phil. Trans. Roy. Soc. London, A180, 187
%
\bibitem[Cutler and Lindblom 1987]{cl87}
 Cutler, L. and Lindblom, L. 1987, \apj, 314, 234
%
\bibitem[Friedman and Morsink 1998]{jfs97} 
 Friedman, J. L. and Morsink, S. M. 1998, \apj, 502, 714
%
\bibitem[Greenspan 1964]{g64}
 Greenspan, H. P. 1964, The Theory of Rotating Fluids (Cambridge: Cambridge 
Univ. Press)
%
\bibitem[Ipser and Lindblom 1991]{il91}
 Ipser, J. R. and Lindblom, L. 1991, \apj, 373, 213
%
\bibitem[Kojima 1998]{k98}
 Kojima, Y. 1998, \mnras, 293, 49
%
\bibitem[Kojima and Hosonuma 1999]{kh99}
 Kojima, Y. and Hosonuma, M. 1999, ``The r-mode oscillations in relativistic
rotating stars'' [gr-qc/9903055]
%
\bibitem[Kokkotas and Stergioulas 1998]{ks98}
 Kokkotas, K. and Stergioulas, N. 1998, \aap, 341, 110
%
\bibitem[Lee 1993]{l93}
 Lee, U. 1993, \apj, 405, 359
%
\bibitem[Lee and Saio 1986]{ls86}
 Lee, U. and Saio, H. 1986, \mnras, 221, 365
%
\bibitem[Lee et al.\ 1992]{lsvh92}
 Lee, U., Strohmayer, T. D. and Van Horn, H. M. 1992, \apj, 397, 647
%
\bibitem[Lindblom and Ipser 1998]{li98}
 Lindblom, L. and Ipser, J. R. 1998, \prd, 59, 044009
%
\bibitem[Lindblom et al.\ 1999]{lmo99}
 Lindblom, L., Mendell, G. and Owen, B. J. 1999, ``Second-order
 rotational effects on the r-modes of neutron stars'' [gr-qc/9902052]
%
\bibitem[Lindblom et al.\ 1998]{lom98}
 Lindblom, L., Owen, B. J. and Morsink, S. M. 1998, \prl, 80, 4843
%
\bibitem[Lockitch and Friedman\ 1998]{lf98}
 Lockitch, K. H. and Friedman J. L. 1998, ``Where are the r-modes of 
isentropic stars ?'' [gr-qc/9812019]
%
\bibitem[Owen et al.\ 1998]{o98}
 Owen, B. J., Lindblom, L., Cutler, C., Schutz, B. F., Vecchio, A. and
 Andersson, N. 1998, \prd, 58, 084020
%
\bibitem[Papalouizou and Pringle 1978]{pp78}
 Papalouizou, J. and Pringle, J. E. 1978, \mnras, 182, 423
%
\bibitem[Press et al. 1992]{pr92}
 Press, W. H., Teukolsky, S. A., Vetterling, W. T., and Flannery, B. P. 
1992, Numerical Recipes: The Art of Scientific Computing, Second Edition 
(Cambridge: Cambridge Univ. Press)
%
\bibitem[Provost et al.\ 1981]{pea81}
 Provost, J., Berthomieu, G. and Rocca, A. 1981, \aap, 94, 126
%
\bibitem[Regge and Wheeler 1957]{rw57}
 Regge, T. and Wheeler, J. A. 1957, Phys. Rev., 108, 1063
%
\bibitem[Sawyer 1989]{s89}
 Sawyer, R. F. 1989, \prd, 39, 3804
%
\bibitem[Saio 1982]{s82}
 Saio, H. 1982, \apj, 256, 717
%
\bibitem[Tassoul 1978]{t78}
 Tassoul, J.-L. 1978, Theory of Rotating Stars (Princeton: Princeton Univ.
Press)
%
\bibitem[Thorne 1980]{th80}
 Thorne, K. S. 1980, Rev. Mod. Phys., 52, 299
%
\bibitem[Unno et al. 1989]{u89}
 Unno, W., Osaki, Y., Ando, H., Saio, H., and Shibahashi, H. 1989, 
 Nonradial Oscillations of Stars (Tokyo: Univ. Tokyo Press) 
%
\end{thebibliography}
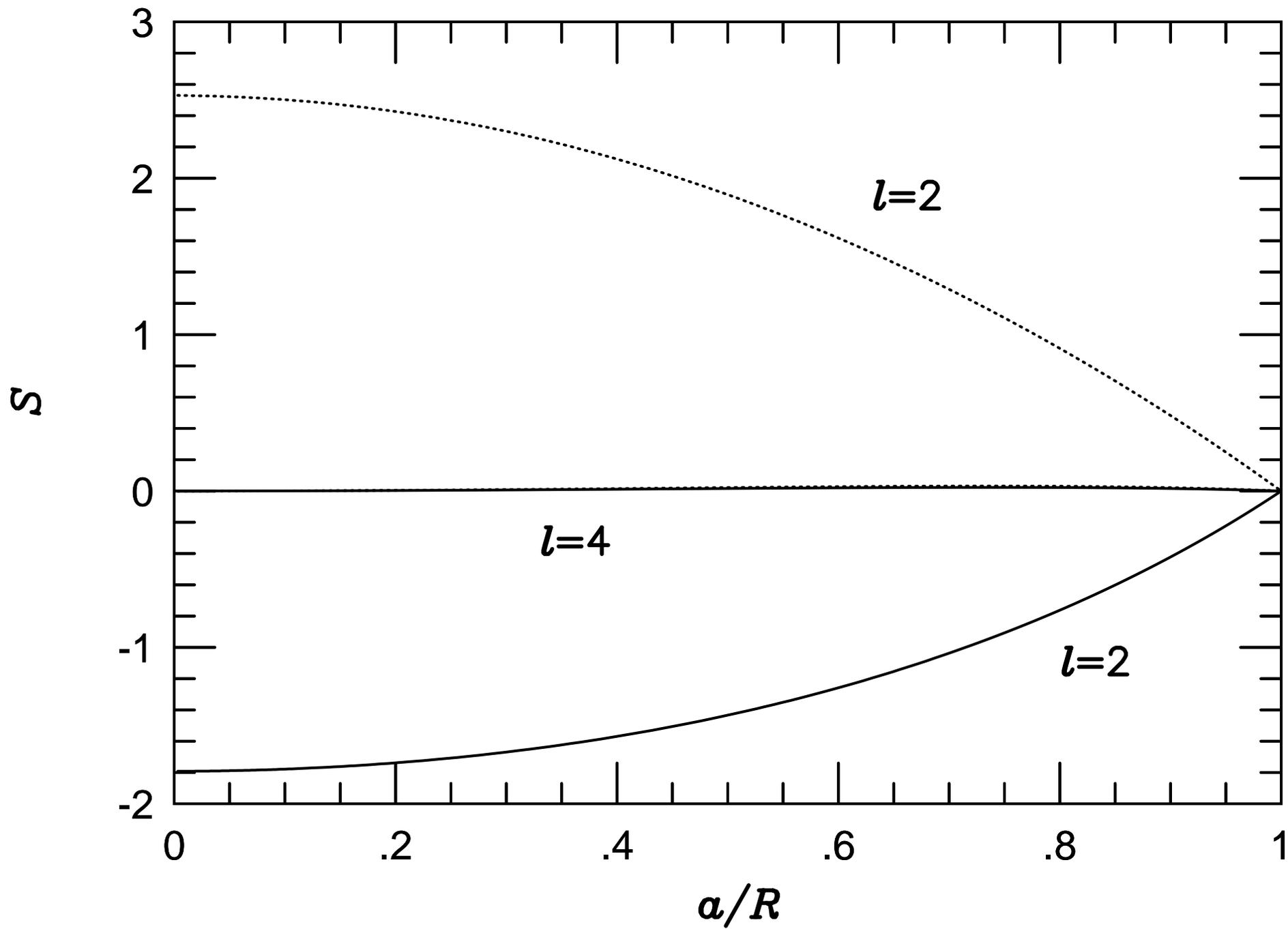
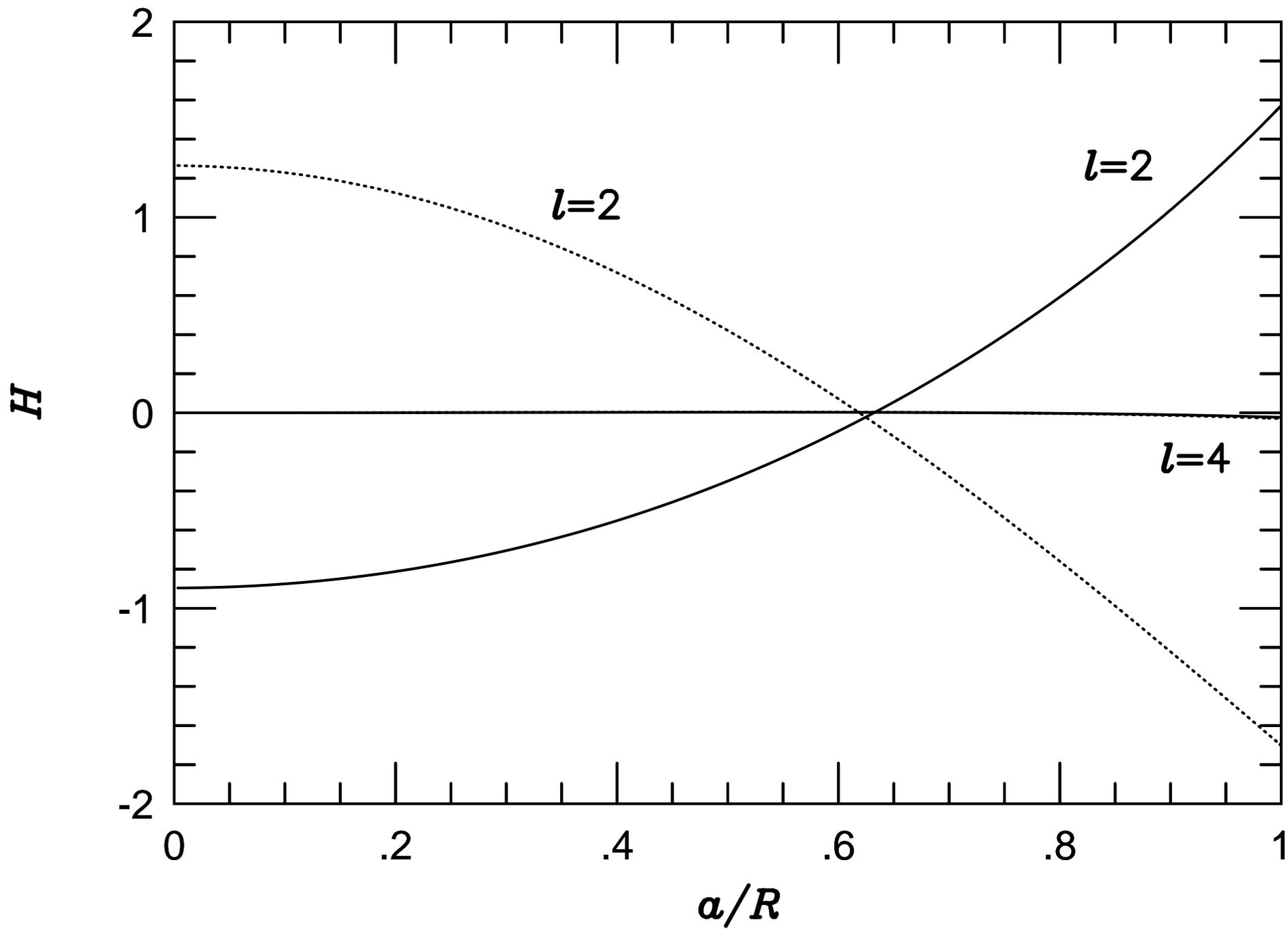
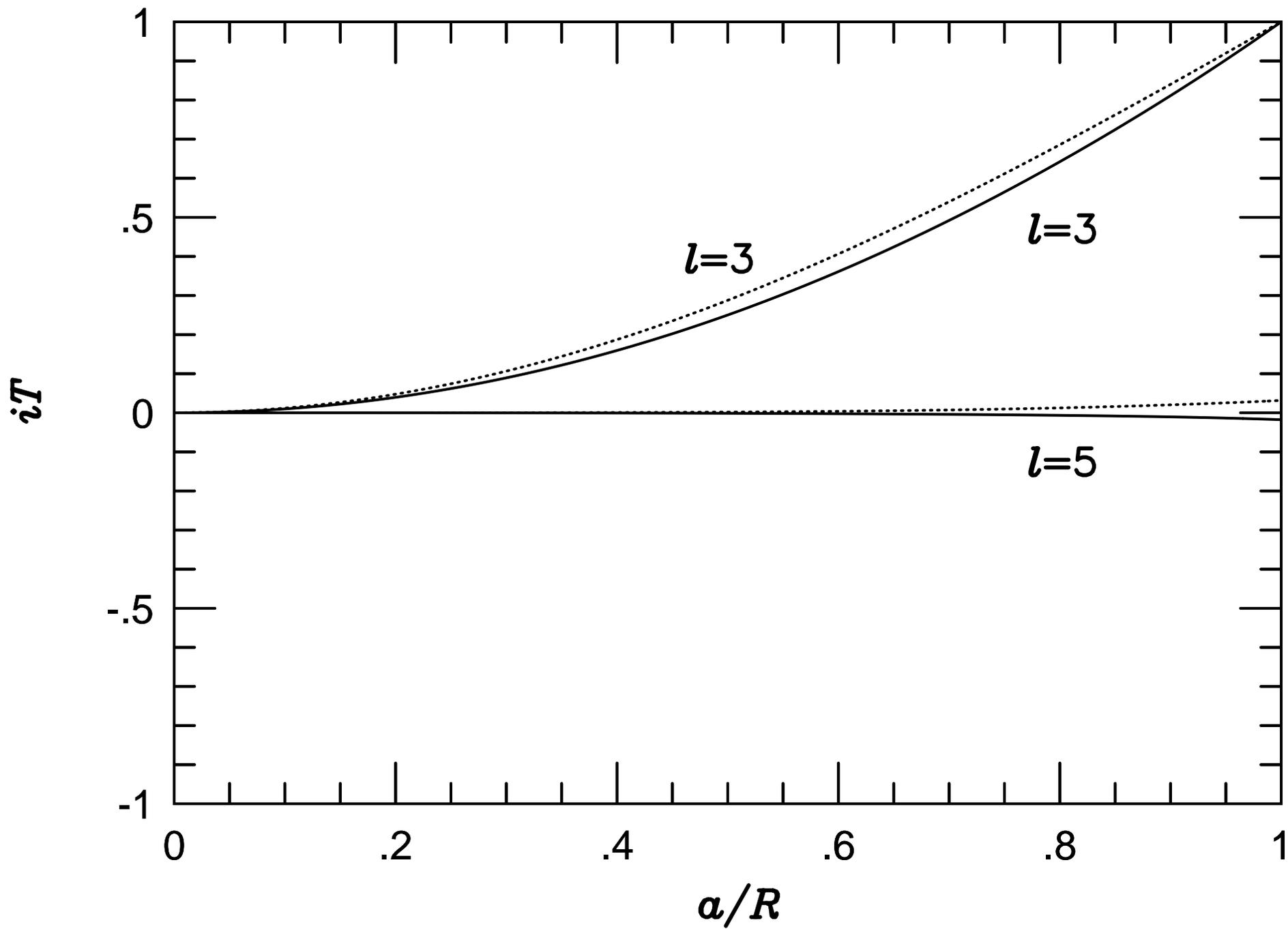
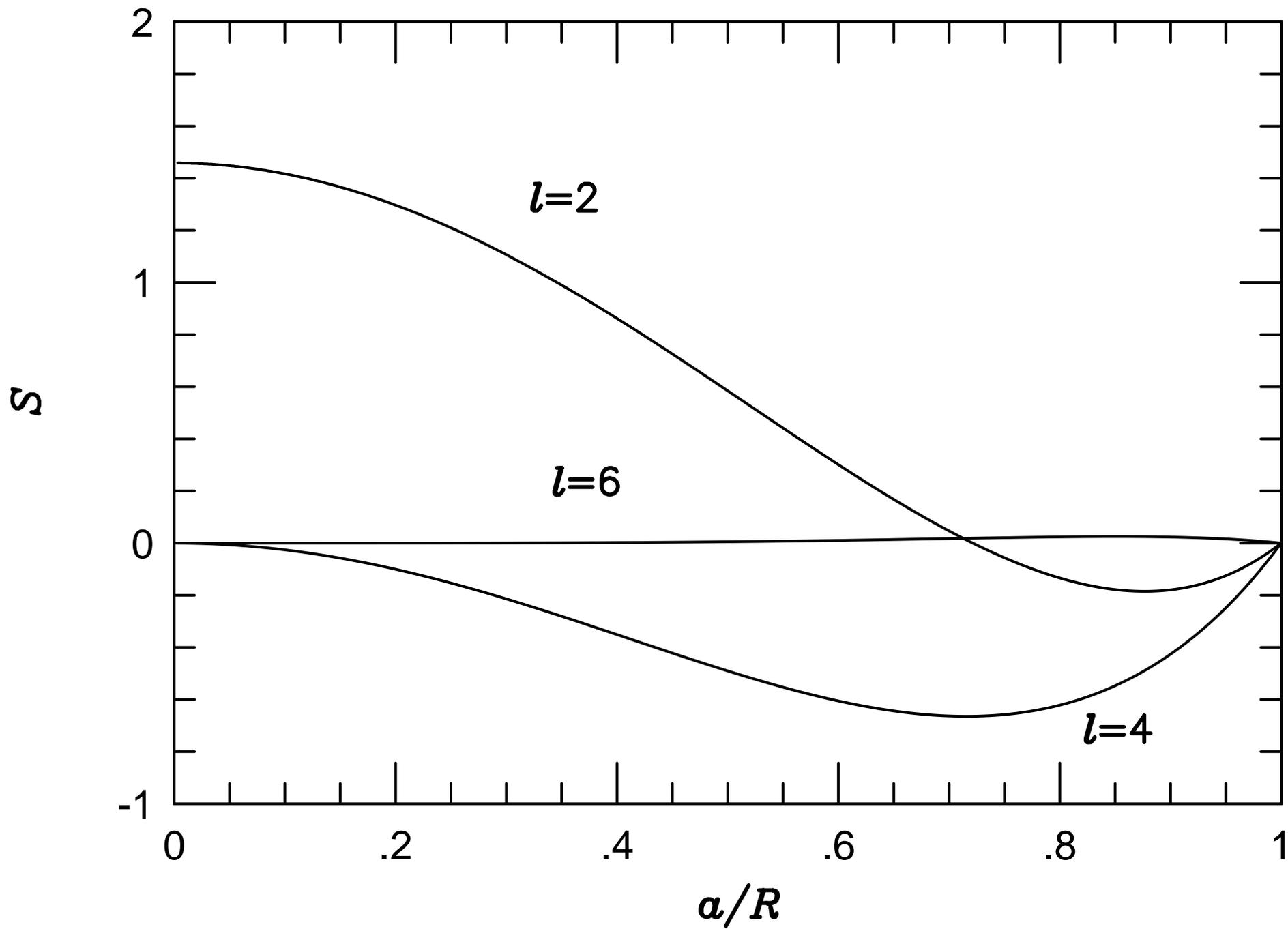
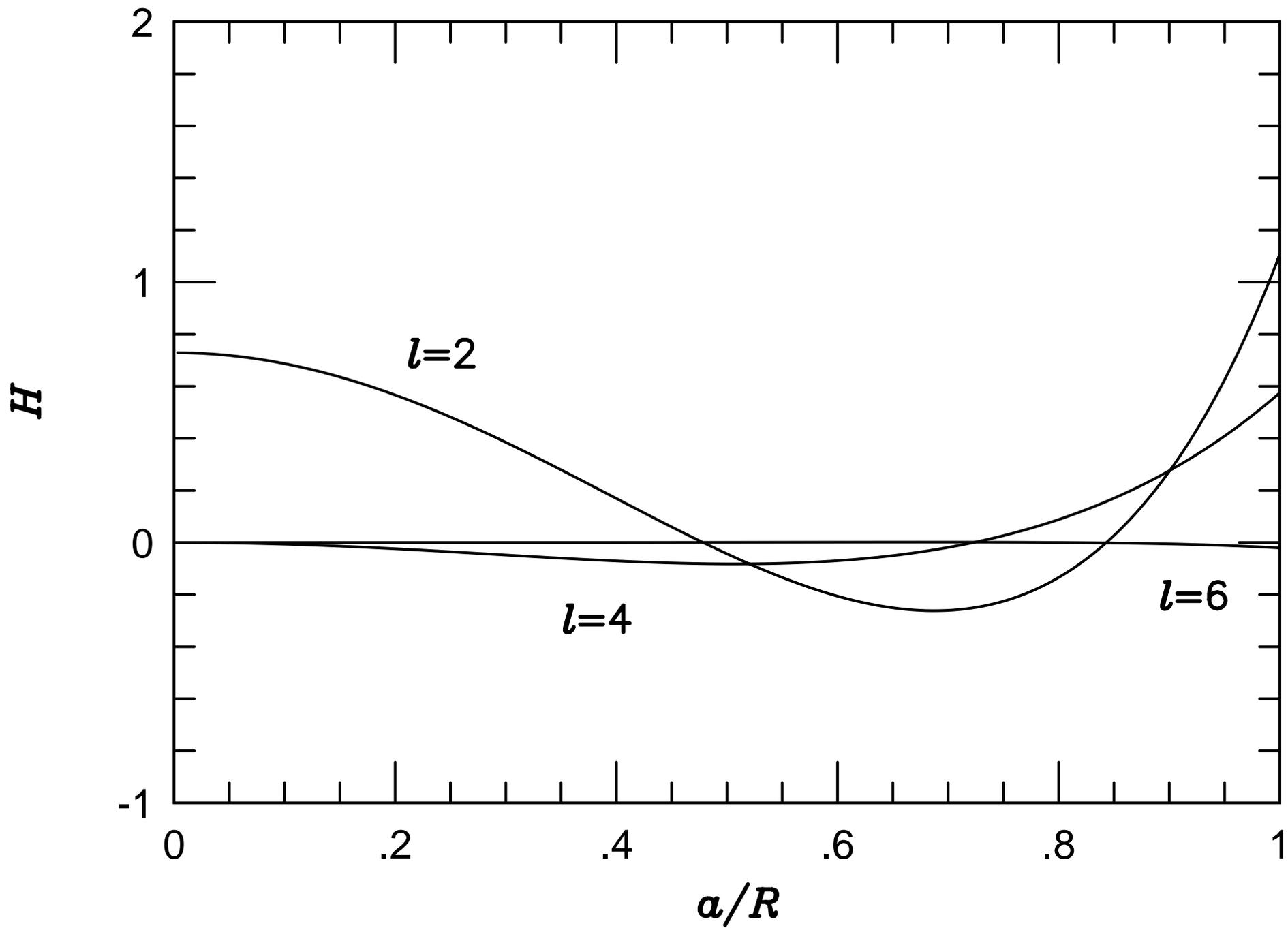
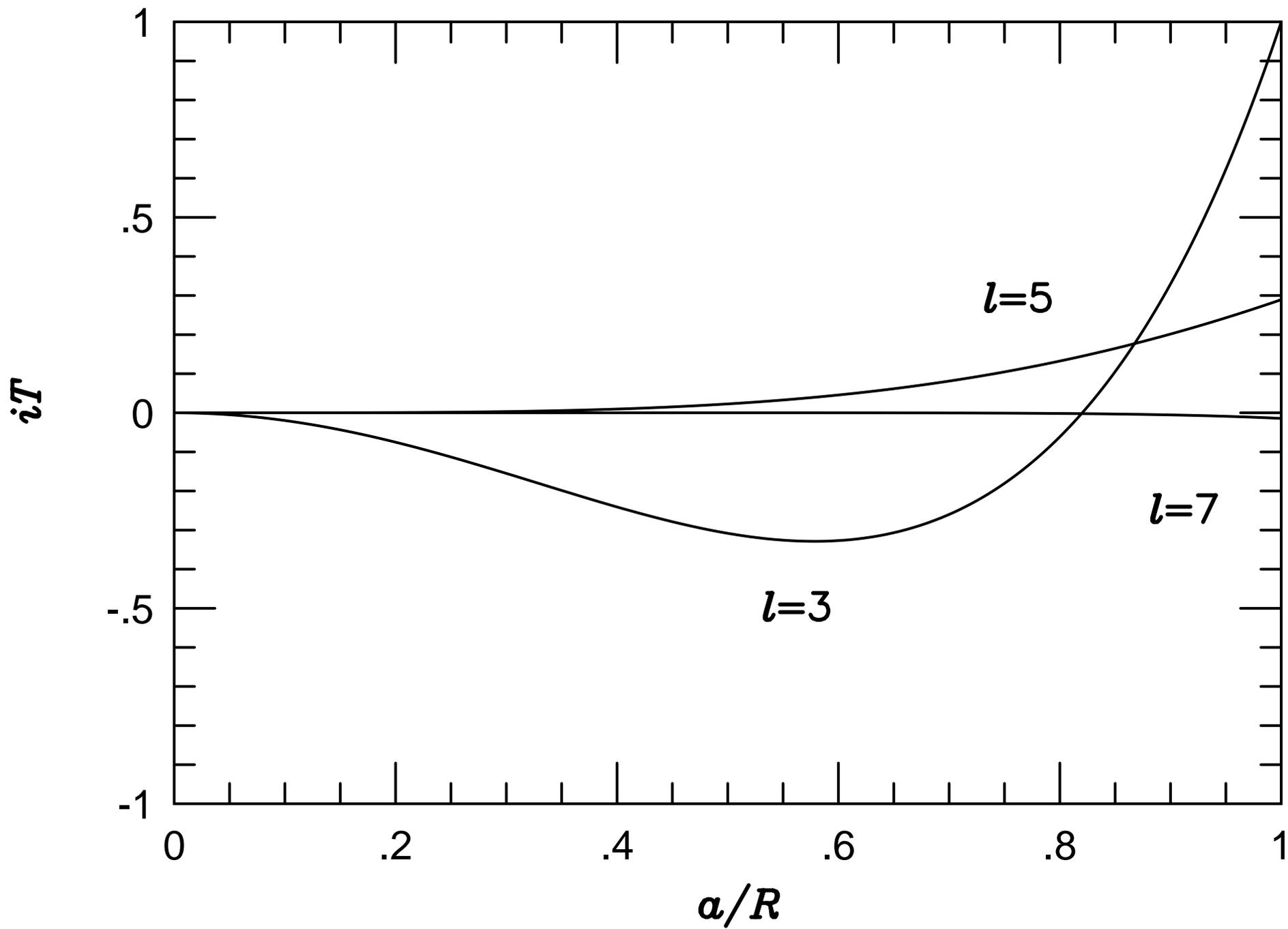
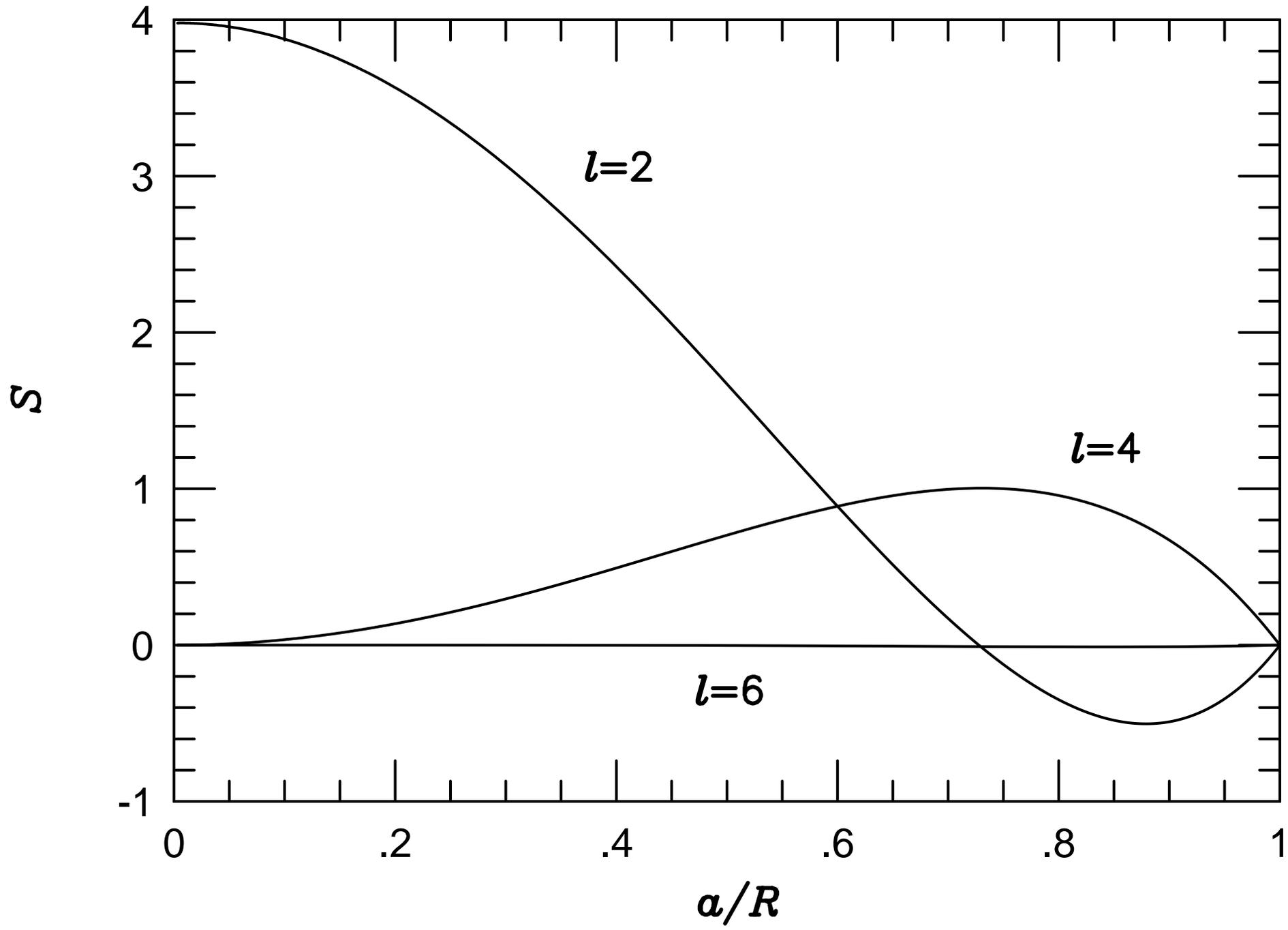
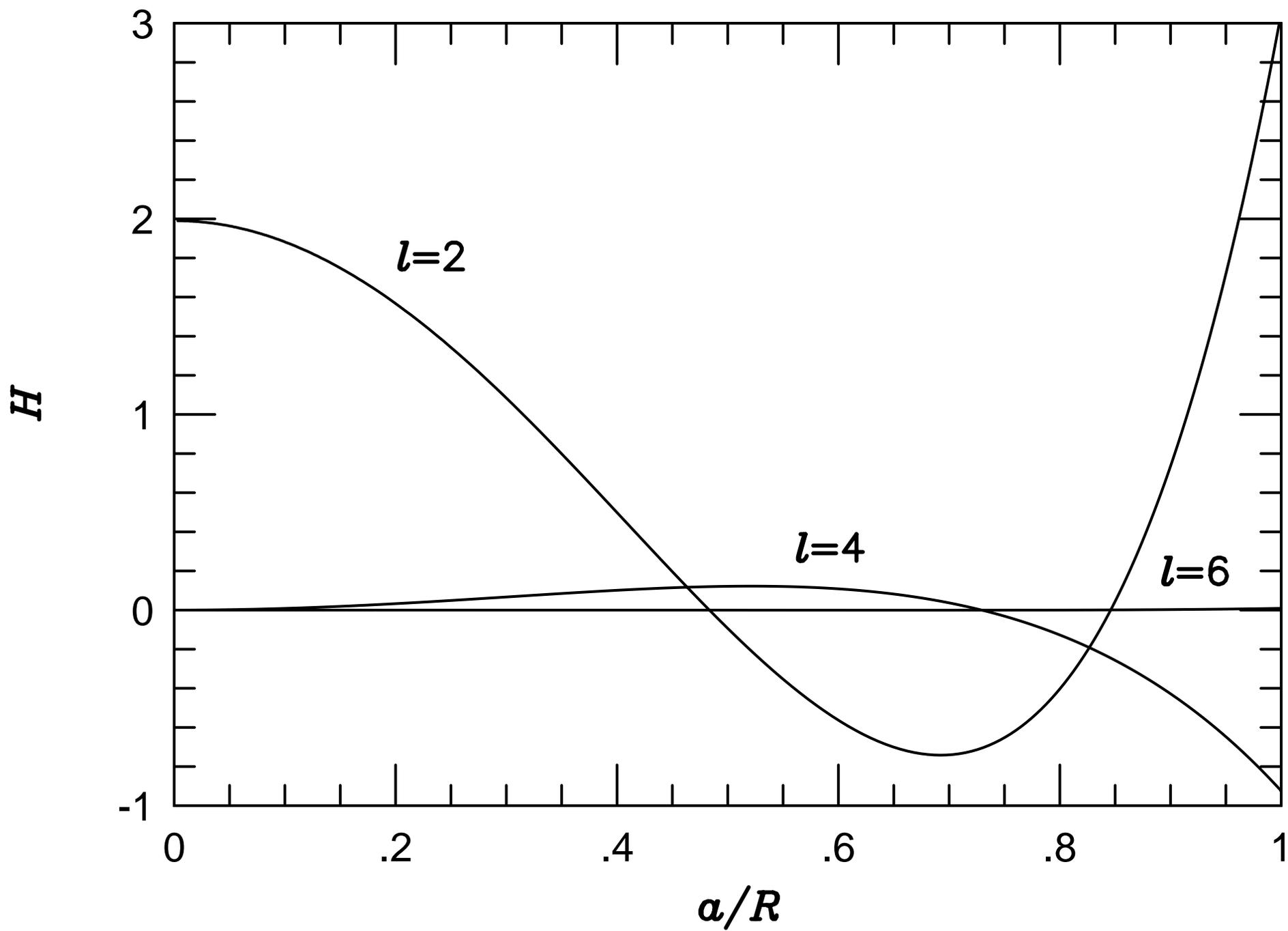
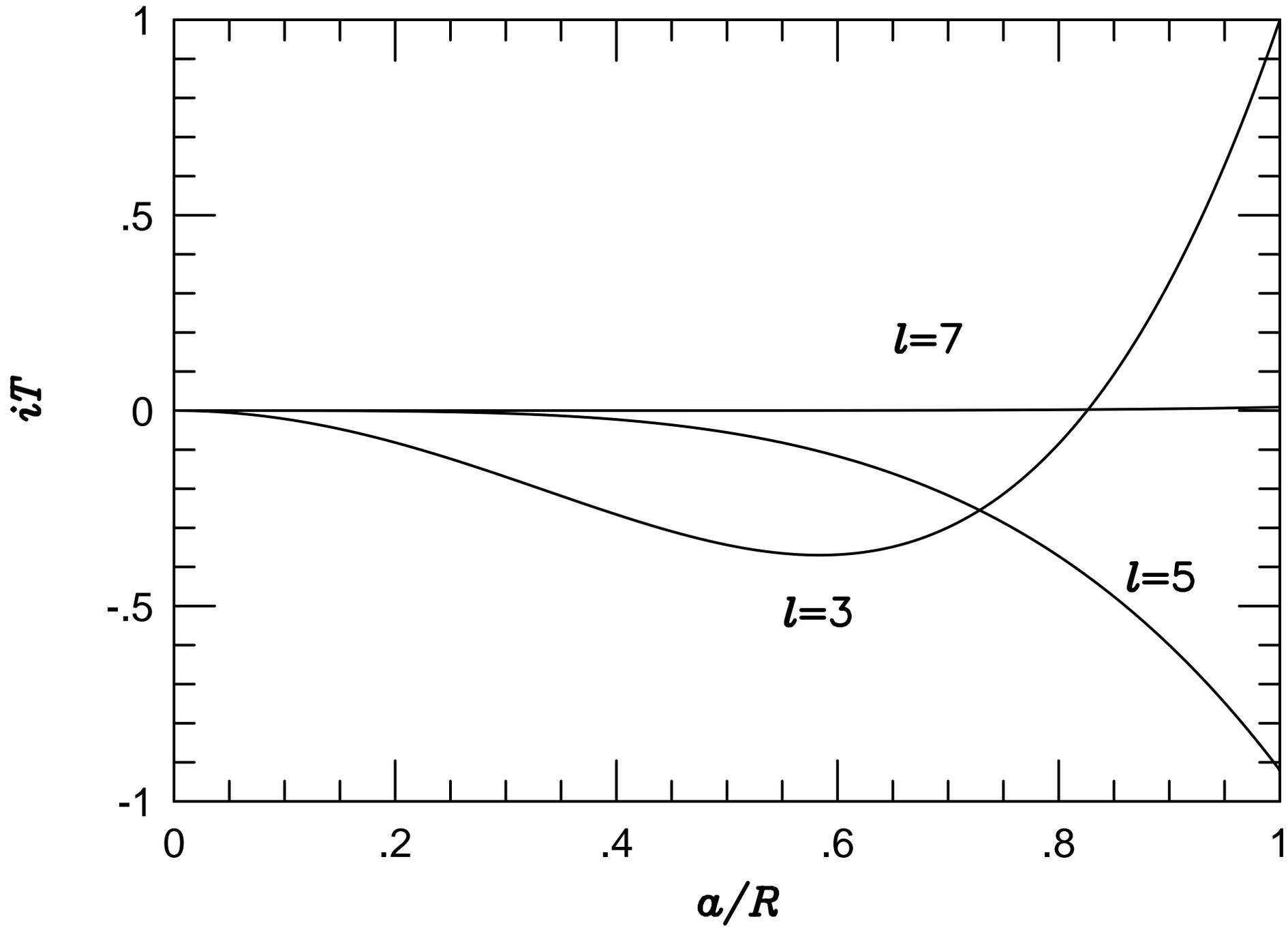
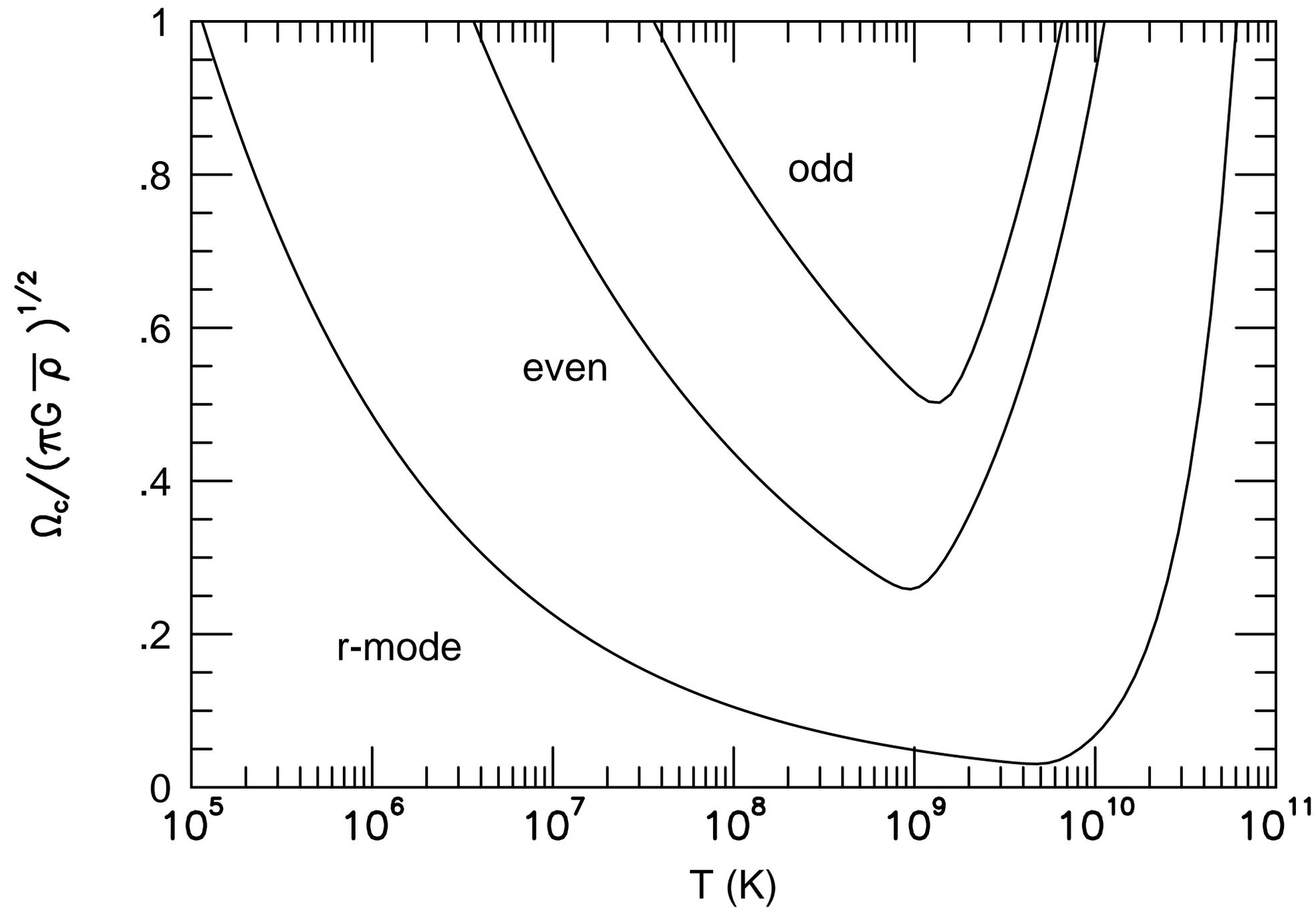
\end{document}